\documentclass[12pt,preprint]{aastex}

\newcommand\pcc{{\,\rm cm}^{-3}}
\newcommand\K{{\;\rm K}}

\newcommand\yr{{\;\rm yr}}

\newcommand\Msun{{\;\rm\,M_\odot}}

\newcommand\kms{{\;\rm km\; s^{-1}}}

\newcommand\pc{{\;\rm\,pc}}

\newcommand\simgt{\lower.5ex\hbox{$\; \buildrel > \over \sim \;$}}
\newcommand\simlt{\lower.5ex\hbox{$\; \buildrel < \over \sim \;$}}

\begin{document}
\title{Protostar formation in supersonic flows: growth and collapse of
  spherical cores}
\author{Hao Gong \& Eve C.\ Ostriker}

\affil{Department of Astronomy, University of Maryland, College Park, MD 20742-2421}
\email{hgong@astro.umd.edu, ostriker@astro.umd.edu}

\begin{abstract}
We present a unified model for molecular core formation and evolution, based
on numerical simulations of converging, supersonic flows. Our model applies to
star formation in GMCs dominated by large-scale turbulence, and contains four
main stages: core building, core collapse, envelope infall, and late accretion.
During the building stage, cores form out of dense, post-shock gas, and become
increasingly centrally stratified as the mass grows over time.  Even for
highly-supersonic converging flows, the dense gas is subsonic, consistent with
observations showing quiescent cores.  When the shock radius defining the core
boundary exceeds $R\approx 4 a (4\pi G \rho_{mean})^{-1/2}$, where $a$ is the
isothermal sound speed, a wave of collapse propagates from the edge to the
center.  During the building and collapse stages, density profiles can be fit
by Bonnor-Ebert profiles with temperature 1.2 - 2.9 times the true value,
similar to many observed cores.  As found previously for initially static
equilibria, outside-in collapse leads to a Larson-Penston density profile
$\rho \approx 8.86 a^2/(4 \pi G r^2)$.  The third stage, consisting of an
inside-out wave of gravitational rarefaction leading to $\rho\propto r^{-3/2}$,
$v\propto r^{-1/2}$, is also similar to that for initially-static spheres, as
originally described by Shu.  We find that the collapse and infall stages
have comparable duration, $\sim t_{ff}$, consistent with estimates for
observed prestellar and protostellar (Class 0/I) cores.  Core building takes
longer, but does not produce high-contrast objects until shortly before
collapse.  The time to reach core collapse, and the core mass at collapse,
decrease with increasing inflow Mach number. For all cases the accretion rate
is $\gg a^3/G$ early on but sharply drops off; the final system mass depends on
the duration of late-stage accretion, set by large-scale conditions in a cloud.
\end{abstract}


\section{Introduction}

Dense molecular cores are the immediate precursors of new stars on
small scales, and understanding how they grow and evolve is
fundamental to the theory of star formation \citep{shu87,mck07}.
Because many elements are involved in core formation, complete
theoretical models have not yet been developed, and it is not yet
clear whether a single dynamical effect dominates the overall process,
or whether several contributing effects have comparable importance.
In one limit that has been studied in some detail, ambient velocities
are negligible, and self-gravitating cores form by the slow diffusion
of partially-ionized gas through strongly-supporting magnetic fields
until a supercritical configuration is reached
(e.g. \citealt{1999osps.conf..305M}).  In another limit, which has
been considered more recently -- but in much less detail at core
scales, magnetic support is negligible, and supersonic turbulence
creates and destroys condensations, with some fraction of this gas
sufficiently dense and long-lived that it can undergo collapse
(e.g. \citealt{mac04}).  As observed clouds are both magnetized and
strongly turbulent, the eventual theory for core formation that is
developed must account for both processes; pioneering work towards this
goal has begun (e.g.
\citealt{2008ApJ...679L..97K,2008arXiv0804.4201N}).  Because of the
technical challenges involved in building comprehensive models and the
need to elucidate the contributing physics, it is important to
develop simplified models in greater detail.  In this contribution, we
consider aspects of core growth and evolution in 
the turbulence-dominated, unmagnetized limit.

Increasingly detailed observations in recent
years provide constraints on theoretical models 
(see e.g. the reviews of \citealt{dif07,war07,2008arXiv0801.4210A}). 
One class of
observations focuses on the density distribution within cores.
One-dimensional (angle-averaged) density profiles 
(e.g. 
\citealt{1994MNRAS.268..276W,shi00,2000A&A...361..555B,
2001ApJ...557..193E,
alv01,kir05,kan05})
generally show a uniform-density center surrounded by a power-law
envelope extending to an outer radius $\sim 0.1
\pc$, which is consistent with the density profile of a static,
isothermal, unmagnetized Bonnor-Ebert (hereafter BE) sphere \citep{bon56,ebe55}.
The interpretation in terms of static equilibrium is problematic,
however, insofar as many cases show center-to-edge
density contrasts exceeding the maximum ratio
($\rho_{c}/\rho_{edge}=14.0$; here $\rho_c$ is the central density) that would be stable
against collapse, and would also require central temperatures greater
than observed values in order to provide support for the total masses
inferred from the integrated continuum emission.  
In addition, cores are generally not isolated; rather than being
surrounded by a high-temperature, low-density medium with pressure
matching the core's outer edge, they are surrounded by
moderate-density cold molecular gas representing clumps and filaments 
within larger clouds \citep{2007ARA&A..45..339B}.
The interpretation
of observed density profiles 
as static solutions is
also not unique, in that dynamically-collapsing cores initiated from
near-critical equilibrium show the
same density profiles as (supercritical) 
static solutions (see e.g. \citealt{mye05,kan05},
and below).  Concentrations formed within turbulent flows can also
have density profiles resembling BE spheres 
\citep{2003ApJ...592..188B}.

Velocity information can distinguish between static, oscillating
\citep{2006ApJ...652.1366K,2007ApJ...671.1832B}, and collapsing cores,
and can potentially also help discriminate how these cores formed out
of more diffuse gas.  Dense, low-mass cores generally have subsonic
internal velocity dispersions, whether for isolated cores or for cores
found in clusters
(e.g. 
\citealt{1983ApJ...270..105M,1998ApJ...504..223G,2002ApJ...572..238C,
2007ApJ...668.1042K,
2007A&A...472..519A,2008ApJ...672..410L}).
In cores containing protostars, signatures of infall on small scales
($\sim 0.01-0.1\pc$), believed to be indicative of gravitational
collapse, have been observed via the asymmetry of molecular lines that
trace high-density gas (e.g.
\citealt{zho93,mar97,1997ApJ...484..256G,dif01}).  For starless cores,
inward motions are often evident over both small ($\simlt 0.1\pc$) and 
larger scales ($\sim
0.1-0.4\pc$), sometimes encompassing a whole star-forming complex
(\citealt{lee99,2001ApJS..136..703L,2006ApJ...637..860W}; see also
\citealt{2006A&A...445..979P}).  Small-scale inward motions within
cores are subsonic \citep{2001ApJS..136..703L}, while 
larger-scale motions
can be transonic or supersonic (\citealt{2006ApJ...637..860W} infer
higher velocities in lower-density gas), 
and may be indicative of converging
larger-scale flows in which dense gas builds up in a shock-bounded
stagnation region.

The relative durations of starless and protostellar (i.e. containing an
accreting embedded Class 0 or I object) 
stages of core evolution are determined by
comparing the relative numbers of the two classes of sources in a
given cloud.  Absolute core lifetimes are further obtained by comparison to 
the number of T Tauri stars with measured ages.  Several 
studies using this statistical approach in different clouds have
reached similar conclusions: the durations of the starless and
accreting stages of cores are comparable
\citep{
1986ApJ...307..337B,
1999ApJS..123..233L,
2000MNRAS.311...63J,
kir05,
2007ApJ...656..293J,
2007A&A...468.1009H,
2008arXiv0805.1075E,
2008arXiv0811.1059E}.
Typical starless core lifetimes are estimated at $\sim 2-5 \times
10^5\yr$, amounting to a few times the free-fall time 
\begin{equation}\label{tff_eq}
t_{ff}\equiv \left(\frac{3 \pi}{32 G \bar\rho}  \right)^{1/2}
=1.37 \times 10^5 \yr\, \left(\frac{\bar n_H}{10^5\pcc}\right)^{-1/2}
\end{equation}
 measured at the mean core density $\bar \rho=1.4 m_H n_H$.
With lifetimes considerably below the ambipolar diffusion time
for strong magnetic fields $t_{AD}\approx 10 t_{ff}$ 
(e.g. \citealt{1999osps.conf..305M}), 
this suggests that observed cores are
gravitationally supercritical with respect to the magnetic field.  
This conclusion is also supported by magnetic field Zeeman
observations, indicating that cores have mean mass-to-magnetic-flux ratios
two times the critical value \citep{2008arXiv0802.2253T}.
Since cores are only identified in millimeter and submillimeter
continuum when the $n_H$ exceeds a few $\times 
10^4\pcc$, in principle it is possible 
that an extended period of slow diffusion at lower density 
precedes the observed core stage.  Turbulence accelerates ambipolar
diffusion, however (e.g. \citealt{2002ApJ...567..962Z,
2002ApJ...570..210F,
2004ApJ...609L..83L,2004ApJ...603..165H}), 
so it is also possible that the flux loss needed to reach
a magnetically supercritical state may occur more dynamically, at densities 
below $10^4\pcc$. 


Theoretical modeling of core evolution has a long
history.  Much work has focused on the evolution of unstable
thermally-supported equilibria into collapse (formally resulting in
infinite density at the origin), followed by accretion of the
envelope.  Self-similar solutions for collapse and/or accretion stages
of isothermal spheres were found by \citet{lar69}, \citet{pen69},
\citet{shu77}, and \citet{hun77}; these were later generalized by
\cite{whi85}.  \citet{lar69} and \citet{pen69} (hereafter LP)
independently found self-similar solutions which describe the density
and velocity prior to the instant of protostar formation (defined as
the instant at which the central density become infinite).  In the LP
solution, the radial 
velocity approaches a constant value $-3.28a$ and the density
approaches
\begin{equation}\label{LPprofile}
\rho=8.86\frac{a^2}{4\pi G r^2}
\end{equation}
at the instant of central protostar
formation, with mass inflow rate $\dot M = 29.1\, a^3/G$. 
Here, $a$ is the isothermal sound speed, and the dimensional factor in
the accretion rate is given by
\begin{equation}\label{accr_def}
\frac{ a^3 }{ G }= 1.6\times 10^{-6}\Msun \yr^{-1}\left(\frac{T}{10 K}\right)^{3/2}.
\end{equation}
The analysis of \citet{shu77} showed that for an initial profile that is a
static singular isothermal sphere, $\rho=2 a^2/(4\pi G r^2)$, 
evolution yields an ``inside-out''
infall solution in which a wave of rarefaction propagates outward at
the sound speed.  Inside of the expansion wave, the mass inflow rate
is $\dot M=0.975\, a^3/G$ independent of $r$, and gas accelerates to
free-fall ($v \propto r^{-1/2}$, $\rho \propto r^{-3/2}$).
\citet{hun77} connected and extended the investigations of LP (which
address evolution prior to protostar formation) with that of Shu
(which focuses on the accretion stage).  He showed that self-similar
solutions before and after the point of singularity formation
(i.e. $t=0$) can be smoothly matched.  This allowed the LP solution to
be extended into the accretion phase with similar free-fall behavior
near the origin; \citet{hun77} also found a
sequence of self-similar solutions valid for all time that approach
the \citet{shu77} expansion wave solution.

Many numerical simulations of isothermal collapse have shown that the
density in the core approaches a $\rho\propto r^{-2}$ profile at the point of protostar 
formation, regardless of how collapse is initiated 
\citep{bod68,lar69,pen69,hun77,fos93, ogi99,hen03,mot03,vor05,gom07}.  These
simulations include initiation from a static configuration that is
unstable, and initiation from static, stable configurations that are
subjected to transient compression, either from enhanced external 
pressure or a
converging velocity field.  Another feature common to the
results from simulations is that the collapse generally begins on the
outside, with the infalling region propagating inward as the central
density increases.  At the time of singularity formation, the central
velocity has been found to be comparable to the value $-3.3a$ derived
by LP, with the density normalization also similar to the LP
result \citep{fos93,ogi99}.  
Following the instant of protostar formation, the evolution
of the mass accretion rate over time depends strongly on the initial
conditions, however.

Simulations with triggered core collapse \citep{hen03,mot03} are
motivated by the fact that star-forming regions are highly dynamic,
such that external compression may significantly affect core internal
evolution, and enhance the accretion rate by raising the central
density. Triggering events may be associated with high mass star
formation, but even without these highly energetic events, the
large-scale turbulence that pervades giant molecular clouds (GMCs) can
compress initially-quiescent cores.  Taking this idea one step
further, it is interesting to consider not just the core collapse
process, but also the core formation process, in a strongly turbulent
medium.  \citet{gom07} conducted one such study, considering
how an impulsive converging velocity field can create gravitationally
bound, centrally-concentrated cores.  Core formation induced by
supersonic turbulence has also been studied in a number of numerical
simulations that focus on the large scales, with much of the emphasis
on determining the distribution of core masses for comparison to
observed core mass functions and the stellar IMF (see e.g. the review
of \cite{mck07}).  However, these studies have not had sufficient
resolution to investigate the internal properties of the cores that
form.  If the mass of a core is built up over time as the post-shock
product of colliding supersonic flows, what is the detailed evolution
leading up to collapse, and during the accretion phase?

In this paper, we initiate a study of dynamically-induced core
formation and collapse by considering perhaps the simplest possible
situation: a supersonic, converging, spherical flow.  Our initial
conditions are a uniform low-density medium with no stratification.
A dense core forms inside a spherical shock, and over time becomes
stratified as its mass grows and it becomes self-gravitating.  When
the stratification becomes too great, collapse and subsequent
accretion occurs in a similar manner to the case of an
initially-unstable static equilibrium.  We consider cases of varying
inflow Mach number, and with the large-scale inflow either steady over
all time, or shut off after an interval.  

The condition of spherical inflow that we adopt for this first study
is, of course, likely to be rare in real clouds.  As the main purpose
of this study is to take the first step towards unified models of
core formation and collapse in dynamic environments, however, we 
consider one-dimensional solutions the natural place to start.  We
shall show that many features consistent with observed cores are evident
even in these idealized models, suggesting that they are generic to
dynamic core formation scenarios.  The present set of simulations, in
addition to enabling identification of characteristic evolutionary
stages, also serve as a useful
reference point for more realistic but more complex simulations.  More
typical than a converging spherical flow would be a converging planar
flow, which yields a dense post-shock stagnation layer in which
self-gravitating cores can form.  Numerical studies that we have begun
for supersonic planar inflows show results for core building and
collapse that are qualitative similar to the present results for
supersonic spherical inflows.

The plan of this paper is as follows:  In \S 2 we present the
governing equations and describe our problem specification and 
numerical method.  Section 3 briefly describes results of collapse initiated 
from static configurations, demonstrating that we reproduce prior
results, and providing a baseline for comparison to our models of
dynamic formation and collapse. Section 4 presents numerical
results for our converging-flow simulations, covering 
the stage of core formation and evolution up to the point
of singularity formation in \S 4.1 and the subsequent stages in 
stages in \S 4.2.  We introduce a breakdown into new physically-defined
stages in \S 4.3, and quantify the evolution of accretion rates in \S
4.4. Section 5 summarizes our new results and discusses our findings in
the context of previous theory and observations.

\section{Governing Equations and Numerical Methods}

The equations of motion for a spherically symmetric flow take the form:
\begin{equation}\label{cont_eq}  
\frac{\partial \rho}{\partial t} + \frac{1}{r^{2}} \frac{\partial
   (r^{2}\rho v)}{\partial r} =0, 
\end{equation}
\begin{equation}\label{mom_eq}
\frac{\partial v}{\partial t}+v\frac{\partial v}{\partial r}=
   -\frac{1}{\rho}\frac{\partial P}{\partial r}-\frac{GM(r)}{r^{2}}, 
\end{equation}
where $M$ is the mass within radius $r$ defined by $d M = 4\pi r^{2}
\rho dr$, $v$ is the radial velocity, $P$ is the gas pressure, and $\rho$
is the density.  For prestellar collapse, an isothermal equation of
state $P=a^{2}\rho$ is often used, 
because cooling is so efficient that the gas remains
at nearly constant temperature during the gravitational collapse 
\citep{lar69,nak98}. We adopt an isothermal equation of state.

For ease of comparison with previous work, we introduce the
following dimensionless variables:
\begin{equation}\label{tau_def}
\tau \equiv t/t_0,
\end{equation}
\begin{equation}\label{xi_def}
\xi \equiv r/r_0,
\end{equation}
\begin{equation}\label{D_def}
D \equiv \rho/\rho_{0},
\end{equation}
\begin{equation}\label{u_def}
u\equiv v/a, 
\end{equation}
\begin{equation}\label{M_def}
m\equiv M/M_0.
\end{equation}
Here $\rho_{0}$ is a fiducial density representing the volume-averaged ambient
density in a cloud on large scales, which we shall later set to
the uniform density of the converging flow. 
The unit of velocity is the isothermal sound speed 
\begin{equation}\label{isosp_def}
a=0.19 \kms \left(\frac{T}{10\K}\right)^{1/2},
\end{equation}
the unit of time is related to the free-fall time at the fiducial 
density by
\begin{equation}\label{t0_def}
t_0\equiv \frac{1}{(4 \pi G \rho_{0})^{1/2}}=0.52\, t_{ff}=
2.3\times 10^6 \yr\, \left(\frac{n_H}{10^2\pcc}\right)^{-1/2},
\end{equation}
the unit of length is related to the Jeans length at the fiducial
density $L_J\equiv a (\pi/G\rho_0)^{1/2}$ by
\begin{equation}\label{r0_def}
r_0\equiv \frac{a}{(4 \pi G \rho_{0})^{1/2}}=\frac{L_J}{2\pi} 
=0.44 \pc \left(\frac{T}{10\K}\right)^{1/2} 
\left(\frac{n_H}{10^2\pcc}\right)^{-1/2},  
\end{equation}
and the corresponding basic unit of mass is 
$\rho_0 r_0^3 = a^3/[4 \pi (4\pi G^3 \rho_0)^{1/2}]$.  
The mass unit adopted in equation (\ref{M_def}) is larger than this by a
factor $4\pi$:
\begin{equation}\label{M0_def}
M_0\equiv \frac{a^3}{(4\pi G^3\rho_0)^{1/2}}
=3.7 
\Msun \left(\frac{T  }{10\K  }  \right)^{3/2}  
\left(\frac{n_{H}}{10^2 \pcc }   \right)^{-1/2}.  
\end{equation}

With the dimensionless
variables, the fluid equations become
\begin{equation}\label{cont_dim_eq}
\frac{\partial D}{\partial \tau} + \frac{1}{\xi^{2}}
  \frac{\partial (\xi^{2} D u)}{\partial \xi} =0, 
\end{equation}
\begin{equation}\label{mom_dim_eq}
\frac{\partial u}{\partial \tau}+u \frac{\partial u}{\partial \xi}
  =-\frac{1}{D}\frac{\partial D}{\partial
  \xi}-\frac{m}{\xi^{2}}, 
\end{equation}
\begin{equation}\label{mass_dim_eq}
m= \int D \xi^{2} d \xi. 
\end{equation}

We solve the 1D hydrodynamic equations (\ref{cont_dim_eq})-(\ref{mass_dim_eq}) 
with the ZEUS-2D
\citep{sto92} code, in spherical symmetry.  
During the evolution to
singularity formation (collapse phase), we adopt an inner reflecting
boundary condition.  For the post-collapse accretion phase, we
implement a sink cell \citep{bos82,ogi99} at the origin when the central
density reaches a reference value.  Subsequently, the inner boundary
condition is changed to outflow, and the value of the central point
mass is tracked via the integrated flow off the grid, with
$\dot M_{ctr} =(a^3/G) D_{in} u_{in} \xi_{in}^2 $.  
The sink cell is only implemented after the inflow in the central region become
supersonic, so that information from the inner boundary 
cannot propagate into the remainder of the grid.

\section{Evolution of Initially-Static Cores}

As discussed in \S 1, many previous numerical simulations of core
evolution have adopted static initial conditions and a fixed total
mass (e.g., \citealt{hun77,fos93,ogi99,vor05}).  The adopted initial
density profile shapes are consistent with (or similar to) a
hydrostatic equilibrium, i.e. a BE sphere, with the initial density
perturbed above the value that can be supported by the internal
pressure in order to initiate collapse.  These fixed-mass simulations
adopt a prescribed external pressure at a low density with an
effectively Lagrangian outer boundary \citep{hun77}, or else a fixed
outer boundary with prescribed external pressure and no mass inflow
\citep{ogi99,vor05}.  \citet{fos93} explored both types of boundary
conditions, and found very similar results for a given initial cloud
density profile.  They concluded that the evolution for the fixed-mass
case is insensitive to the outer boundary condition for
initially-unstable equilibria.  This is consistent with the argument
of \citet{bod68} that the outer boundary condition does not affect
evolution up to collapse as long as the free-fall time
(eq. \ref{tff_eq}) is shorter than the cloud crossing time.  The ratio
of the free-fall time at the mean density to the sound crossing time
$r_{max}/a$ over the radius of a BE sphere is $\pi/(8\xi \frac{d
\Psi}{d \xi})^{1/2}$, which approaches $\pi/4 \sim 0.785$ as $\xi$
approaches infinity.  Here, $\Psi=\Phi/a^2$, the dimensionless
gravitational potential.  For the critical case,
the free-fall time is 0.71 times the sound crossing time over the radius.

For comparison to previous work, we consider collapse of an
initially-static BE sphere.  For initial conditions, we adopt a critical  
BE sphere, i.e. the outer boundary of the grid is at radius
\begin{equation}\label{radiusBE}
R_{BE,crit}=6.45\frac{a}{(4\pi G\rho_c)^{1/2}}=1.72\frac{a}{(4\pi G\rho_{edge})^{1/2}}
=2.70\frac{a}{(4\pi G\rho_{mean})^{1/2}}
\end{equation} 
corresponding to dimensionless outer radius $\xi_e= \xi_{crit} = 6.45(\rho_0/\rho_c)^{1/2}$. 
Here, $\rho_{mean}$ is the total core mass divided by its volume.
The mass of the critical BE sphere is
\begin{equation}\label{massBE}
M_{BE,crit}= 1.18 \frac{a^4  }{(G^3 P_{edge})^{1/2}}
=4.18 \frac{a^3  }{(4\pi G^3 \rho_{edge})^{1/2}}
=1.5 \Msun \left(\frac{T  }{10\K  }  \right)^{3/2}  
\left(\frac{n_{H,edge}}{10^4 \pcc }   \right)^{-1/2}.  
\end{equation}
To initiate collapse, density is perturbed above the
equilibrium value by $10\%$.  Our outer boundary condition is at a
fixed pressure, with no inflow.
The temporal evolution of the accretion rate for this model is
shown in Figure \ref{fig:mfl_pre}.

Features similar to those outlined by \citet{vor05} are
observed in our simulation. The accretion rate peaks at a value
approaching the LP prediction at the
instant when the central density (formally) becomes infinite, and
steeply decreases thereafter.
\citet{vor05} find that the decline in the accretion rate after 
singularity formation (unlike the increase in the accretion
rate in Hunter's self-similar extension of the LP solution) can be
attributed to the variation of velocity with radius in the sphere as
it evolves toward collapse.
We also considered cases of much larger initial static spheres, 
with outer boundaries $\xi_e = 5\, \xi_{crit}$ corresponding to highly-unstable BE
configurations.  For these cases, the accretion rate decreases until
it reaches a plateau at $\dot M = 1.45\,a^3/G$, consistent with the value
reported in \citet{shu77} (when the density is 10\% greater than
for hydrostatic equilibrium), and then further declines to zero after a
rarefaction wave propagates inward from the outer boundary to reach
the center.

Similar to the results of previous simulations, we find that the first
collapse is ``outside-in,'' with velocities initially nonzero only in
the outer parts where the imbalance between gravity and pressure is
largest.\footnote{For cases where the initial sphere is larger than the
  critical BE sphere, collapse begins at radii near $\xi_{crit}$, as
  was previously shown by \citet{fos93}.}
This is because the inner portions of the sphere, at
$\xi<\xi_{crit}$, initially are equivalent to {\it stable} BE
solutions.  In all of our models initiated from static spheres, the
density profile approaches the LP self-similar solution
$D=8.86\xi^{-2}$, and the velocity in the inner region approaches
$-3.28a$, at the moment of singularity formation. Before this time,
the density profile in the central region is flat with a magnitude
that increases over time.  The process can be thought of as a wave of
compression propagating from the outside to the inside, creating
a density distribution in which the ratio of radius to the Jeans
length at the local density is everywhere
constant: $r/L_J(r)\approx \sqrt{8.86}/2\pi\approx 0.47$. The 
singularity represents the instant the
compression wave converges at the center. 

Supersonic inflow velocities
can be achieved without shock formation in the interior of the core
(except at $r=0$) 
because inward acceleration occurs at all radii where the inward 
gravitational force exceeds the outward pressure force.  By
construction, the solutions initiated from static configurations all
have gravity (slightly) exceeding pressure forces everywhere in
the initial state, so that the inward acceleration is nonzero.
After a singularity forms at the center, accretion begins, and the
flow in the interior transitions from the $v=const.$, $\rho\propto
r^{-2}$ LP solution to a free-fall solution, with the accreting region
propagating from the interior to the exterior in a manner similar to
that described by \cite{shu77}.  Thus, while collapse
develops in an ``outside-in'' fashion, accretion develops in an
``inside-out'' fashion.  Note that accretion in any
centrally-concentrated configuration should work its way outward from
the center, because gravitational collapse times decline outwards
$\propto \rho^{-1/2}$, which is $t_{ff}\propto r$ for an inverse-square
density profile (true for either the LP profile or the singular
isothermal sphere).

\section{Converging-Flow Model Results}

In this section, we present the results of our simulations of core
formation and evolution for the case of converging spherical,
supersonic flows.  For these simulations, the outer boundary condition
at $\xi_{max}$ is maintained at a constant density and inflow
velocity.  The inflow velocity is characterized by the Mach number 
relative to the isothermal sound speed,
$M_a\equiv v_{in}/a$.  We vary $M_a$ over the range from 1.05 to 7.
For some models (see below), we suppress inflow at the outer radius
after collapse occurs to test how the late-time accretion rate is altered.
The initial conditions consist of uniform (low) density, and uniform
inflow velocity equal to the value at the outer boundary.
The size of the grid, in terms of the reference length scale
given in equation (\ref{r0_def}), is $2.51327 r_0$, which amounts to a fraction
0.4 of the Jeans length at the initial density. Note that the radius
of a critical BE sphere at the same external pressure (so that 
$\rho_{edge}=\rho_0$) would be 
$R_{BE,crit}=6.45(\rho_{edge}/\rho_c)^{1/2} r_0=1.72 r_0$. That is,
a critical BE sphere confined by the same ambient pressure would be
able to fit within our simulation domain, with plenty of room to spare.
The size of zones in the radial direction has a constant logarithmic
increment, i.e., $\Delta r_{i+1} = (1+\alpha) \Delta r_i$, for some
$\alpha > 0$, such that 
$r_i= r_1 + \alpha^{-1}[(1+\alpha)^{i-1}-1]\Delta r_1$
and $\Delta r_1/(r_{\rm max}-r_{\rm min})= \alpha [(1+\alpha)^{N-1} -1]^{-1}$.
For all the converging flow simulations in this paper,
$\alpha$ is set to $0.009$; 605 and 597 grids are used during the
collapse phase and accretion phase respectively.

\subsection{Core Formation and Collapse}

We begin with a description of the core formation process, which is
similar for all of the converging-flow models.  Because of the
reflecting boundary condition at the center and the initial inflow 
velocities, immediately after we initiate the simulation, 
a shock forms at the origin and propagates outward. The
inflowing matter is compressed by the shock.
The shock front divides the converging inflow
into two regions: an inner dense post-shock region 
and an outer low-density region of supersonic inflow.  
These two regions evolve quasi-independently but are connected by
shock jump conditions. Under competition between gas pressure and
self-gravity, the inner region contracts slowly to begin forming a 
dense core. As self-gravity starts to overwhelm 
gas pressure, the dense core enters the collapse phase.

At the beginning, the inner region is quasi-hydrostatic, with the velocity a
linear function of radius. With negligible density gradient
and self-gravity in the early stages, equation (\ref{mom_dim_eq}) becomes
\begin{equation}
\frac{\partial u}{\partial \tau} + u \frac{\partial u}{\partial \xi} 
\approx 0.  
\end{equation}
This equation is satisfied by 
\begin{equation}\label{u_law}
u = \frac{ \xi}{\tau - \tau_0}; 
\end{equation}
for $|\tau/\tau_0|\ll 1$, $u \approx - \xi/\tau_0$, i.e. the
coefficient of the linear profile is constant in time. The leftmost lower panels 
of Figures \ref{fig:BE_shock_1.05} - \ref{fig:BE_shock_7.0}
show this linear-velocity behavior in the shocked region, at the time 
that equals half of the collapse time $\tau_{coll}$. Throughout this
paper, we define the collapse time $\tau_{coll}$
as the moment
that $\rho_c/\rho_0 = 4\times 10^7$. 
Shown in the leftmost upper panels of Figures 
\ref{fig:BE_shock_1.05} - \ref{fig:BE_shock_7.0}
 are the density profiles; even when
the density profile has nonzero gradients, the velocity in the inner
region is dominated by the linear term.
Notice that the leftmost panels have linear-linear scales.
The velocity over the whole post-shock region is subsonic and negative,
which means the core is slowly contracting. For the $M_a =4, 7$ models
(see Fig. \ref{fig:BE_shock_4.0}, \ref{fig:BE_shock_7.0}),
the inner part of the density and velocity profiles oscillate at the
beginning of simulations. As the shock front propagates outward,
the mass inside the shock increases, and so does self-gravity.
After a period of accumulation lasting about 90\% of the time until
collapse $\tau_{coll}$,
the slowly-contracting dense region starts to be gravitationally unstable.

Note that the density and velocity profiles of the regions outside of the
shock go through a transient evolution after simulations begin. The density
profile evolves from a uniform profile set in the initial condition to a 
$\rho \propto r^{-2}$ profile consistent with supersonic radial 
inflow\footnote{For supersonic radial flow, 
$v \sim const.$ upstream of the shock combines with the 
steady state mass inflow 
condition $\rho v r^2 = const.$ to yield $\rho \propto r^{-2}$.}; 
the material is also increasingly accelerated by gravity. In Figures 
\ref{fig:BE_shock_1.05} - \ref{fig:BE_shock_7.0}, the inflection feature in
the velocity profiles exterior to the shock corresponds to a
wave propagating inward at a speed equal to the inflow speed plus the sound speed.

Since post-shock velocities are subsonic, gravitational instability is 
expected to develop at a point when the radius
of the inner dense post-shock region becomes comparable to the critical
radius of a BE sphere. This expectation is indeed borne out by our
simulations, which moreover show that the properties of the collapsing
dense inner region are similar as those of collapsing cores initialized 
from hydrostatic BE spheres.
The collapse follows an ``outside-in'' pattern, starting from the shock front.
 The central density increases dramatically and the inflow velocity inside the
shock becomes supersonic. 
The collapse propagates inward and establishes a centrally-concentrated density profile
\citep{lar69,pen69,hun77,shu77}, which approaches the LP density profile 
$D(\xi) = 8.86 \xi^{-2}$, and the velocity approaches $-3.28 a$. 

The three panels on the right of Figures \ref{fig:BE_shock_1.05} - 
\ref{fig:BE_shock_7.0} show the density profile (top) and the
velocity profile (bottom) at three different instants during the
outside-in collapse. The density profiles plotted are normalized by the central 
density and the interior regions are fit by BE sphere density profiles.
\footnote{That is, we match solutions of the density profile inside
the shock to solutions of the hydrostatic equation 
$\frac{d\mathrm{ln}D}{d\xi}=-\frac{m(\xi)}{\xi^2}$
with a fitted temperature $T_{BE}$. These fits are further discussed below.}
The ratio between the fitted BE sphere temperature $T_{BE}$ and the true 
temperature $T_0$ is also noted in the figures. 
The first figure of these three shows profiles at the instant when the radius of the 
post-shock region reaches the critical radius (see eq. \ref{radiusBE})
of a BE sphere at temperature $T_0$ 
with the same central density i.e. $r_{shock} = R_{BE,crit}$.
We define the period after this as the supercritical regime. The center-to-edge
density ratio after this point exceeds $\sim 10$.
The second figure of these three shows the instant when the radius of
the post-shock region reaches twice the critical radius of a BE
sphere. The third figure is at the instant the core collapse $\tau_{coll}$ 
(defined here as $\rho_c/\rho_0 = 4 \times 10^7$). 
The long dashed diagonal line in the third figure shows the LP density profile,
which is very close to the numerical solution.
The time interval between $r_{shock}=R_{BE,crit}$ and collapse for the $M_a =2$ case
in code units is 0.048 (see eq. \ref{t0_def}), which corresponds to 
$1.1\times 10^5 \yr$ if the inflowing ambient medium's density is $n_H = 100 \pcc$.

We note that prior to collapse, the velocities in the dense gas (inside the
shock) remain small. In particular, for $M_a \ge 2$ cases, 
the inflow velocity inside the shock is subsonic throughout the post-shock 
region before the time when $r_{shock} = R_{BE,crit}$.
When $M_a$ is high, Figures \ref{fig:BE_shock_4.0} and \ref{fig:BE_shock_7.0} 
show that the post-shock velocities remain subsonic up to the instant of collapse.

The outward propagation speed of the shock, which from the simulations
is approximately constant (Fig. \ref{fig:shkfrps}) at early time, can 
be obtained using the shock jump conditions. If the shock position is
$\xi_{sh}=u_{sh}\tau$, then provided equation (\ref{u_law}) is
satisfied, the velocity on the downstream side is
\begin{equation}\label{u_d_eq}
u_d=\frac{\tau u_{sh}}{\tau-\tau_0}. 
\end{equation}
Notice that in order for the downstream velocity to be inward and the
shock to be propagating away from the origin, we must have $\tau<\tau_0$.
The isothermal 
shock jump conditions, with subscript ``$u$'' denoting upstream
and ``$d$'' downstream values, are
$D_d (u_d- u_{sh}) = D_u (u_u-u_{sh})$ 
and 
$D_d[ 1 + (u_d -u_{sh})^2] = D_u[1+ (u_u-u_{sh})^2]$; 
together these imply 
$D_d/D_u=(u_u-u_{sh})^2$ and $(u_d-u_{sh})(u_u-u_{sh})=1$.
Treating the shock speed as approximately constant so that equation
(\ref{u_d_eq}) holds, we can solve for the shock velocity to obtain
$u_{sh}=\frac{1}{2}(u_u + [ u_u^2 + 4- 4(\tau/\tau_0)]^{1/2})$.
For $\tau/\tau_0\ll 1$, and taking the upstream velocity as $u_u\approx-M_a$
which corresponds to the limit of strongly supersonic inflow, gives 
\begin{equation}\label{eq:ush}
u_{sh} \approx \frac{1}{2}\left[-M_a+ (M_a^2 + 4)^{1/2}\right]
=\frac{2}{M_a+ (M_a^2 + 4)^{1/2}}.
\end{equation}
The shock speed decreases as the Mach number increases, and therefore
from equation (\ref{u_d_eq}) the post-shock flow speed also decreases
as the Mach number increases. Figure \ref{fig:shkfrps} shows the position
of shock front versus time for $M_a =4$. The intercept and the slope are 
based on the best fit of the linear part where $\tau \in$ [0.0, 0.04]. 
The intercept is nearly 0 and the slope 0.3061 is the measured shock speed in units 
of the isothermal sound speed $a$; equation (\ref{eq:ush}) predicts a slightly
smaller value 0.24. The analytical solution (eq. \ref{eq:ush})
for $u_{sh}$ as a function of $M_a$ is plotted as a dotted line
in Figure \ref{fig:u_xi_tau_ori} (labeled as $u_{sh,estimate}$) and the 
shock speeds measured directly from simulations $u_{sh}$ are plotted as 
asterisks. The analytical approximation is about 15-28\%
below the measured value from the simulations as $M_a$ ranges from 1.05-7.0.

Using the constant-shock-speed approximation, the immediate post-shock
density can also be obtained in terms of the pre-shock density, in the
limit $\tau/\tau_0\ll 1$ and using $u_u \approx -M_a$ as:
\begin{equation}\label{eq:rho_d_u}
\frac{D_d}{D_u}=
\frac{1}{4}\left[M_a + (M_a^2+4)^{1/2}\right]^2,
\end{equation}
which for $M_a\gg 1$ is $D_d/D_u \approx M_a^2$.  Because of the
radial convergence of the inflow in the simulations, $D_u$ varies; 
it is initially equal to $1$, but after an initial transient, in the
highly-supersonic limit $D_u$ would approach $(\xi_{outer}/\xi_{sh})^2$
because of mass conservation.

It is interesting to investigate how the state of the core when it
collapses depends on Mach number.  Figure \ref{fig:rho_v_collapse} shows the 
density profiles and the velocity profiles of the simulated cores at the 
instant of collapse $\tau_{coll}$, for different Mach numbers.
Firstly, it is clear that all the density profiles approach
the LP solution, especially for low Mach number cases. The inflow velocity 
inside the shock is supersonic and does not strongly vary over the dense core region.
The smaller the Mach number is, the closer the inflow speed is to the LP 
result, $v=-3.3 a$. 
Secondly, the radius of the core at the instant of collapse decreases 
as Mach number increases. We plot this dependence in Figure \ref{fig:u_xi_tau_ori}
with diamonds. 

We quantitatively compare the basic core properties in Figure \ref{fig:u_xi_tau_ori},
which shows the collapse time $\tau_{coll}$ and the shock radius $\xi_{coll}$ at
time $\tau_{coll}$, both as function of $M_a$. As noted above, $\xi_{coll}$ decreases
with increasing $M_a$; the same is true for $\tau_{coll}$. We also recall that $u_{sh}$
decreases with $M_a$. Interestingly, while $\xi_{coll}, u_{sh}$, and $\tau_{coll}$ 
all decrease with $M_a$, the ratio $\xi_{coll}/(u_{sh}\tau_{coll})$ is nearly constant 
with $M_a$: it ranges only from 0.34 to 0.42 (see Fig. \ref{fig:u_xi_tau_ori}). 
This result is potentially useful for 
empirical estimates of core lifetimes, since the inflow velocity $M_a$, 
the isothermal sound speed $a$ and the radius of dense core are all in principle
measurable. If the ratio $\xi_{coll}/(u_{sh}\tau_{coll})$ is taken as a constant
$\approx 0.4$, and the shock speed is estimated via equation (\ref{eq:ush}), then
the life time of cores up to the point of collapse is given in dimensional form
by $t_{coll} \approx 1.3 R_{coll} [M_a+(M_a^2+4)^{1/2}]/a$. In practice, it may
be difficult to measure $M_a$ outside of a core, because the density is much lower
than that of the core, and it is difficult to isolate the immediate environment of 
the core from foreground and background gas. This result is still useful in a 
statistical sense, however, using the mean Mach number of the turbulent flow in 
a cloud.

Another direct observable is the core density, so it is interesting to test how
the values of the collapse time and radius depend on the mean density in the core
at the time of collapse. Diamonds in the top panel of Figure \ref{fig:u_xi_tau_nor} 
show the core radius in units of $a(4\pi G \rho_{mean})^{-1/2}$, 
which is $\xi_{coll}(\rho_{mean}/\rho_0)^{1/2}$, as a function of $M_a$.
We can see $\xi_{coll}(\rho_{mean}/\rho_0)^{1/2}$ is nearly constant, 
ranging from 4.58 to 3.42 as Mach number increases from 1.05 to 7. 
Taking this as approximately constant, and taking the measured core mean density, 
the predicted
size of core at the time of collapse is $R_{coll}\approx 4a(4\pi G \rho_{mean})^{-1/2}$. 
Note that this radius is $\sim 50\%$ 
larger than the critial BE radius for the same temperature
 (see eq. \ref{radiusBE}).
Since the post-shock density increases relative to the upstream density approximately
as $D_d/D_u \sim M_a^2$ (see eq. \ref{eq:rho_d_u}), it is also interesting to test how 
$\xi_{coll}M_a$ depends on Mach number. In fact this quantity decreases with $M_a$,
as seen in Figure \ref{fig:u_xi_tau_nor}.

To express the core collapse time in terms of observables, we normalize the collapse 
time using the mean core density. This quantity $t_{coll}(4\pi G\rho_{mean})^{1/2}=
\tau_{coll}(\rho_{mean}/\rho_0)^{1/2}$, is plotted in Figure \ref{fig:u_xi_tau_nor} as
a function of $M_a$. For reference, the core collapse time normalized using $M_a$ is 
also plotted in Fig. \ref{fig:u_xi_tau_nor}. The free-fall time $t_{ff}$ for a uniform
sphere in units 
of $(4\pi G \rho_{mean})^{-1/2}$ is $1.92$, so that we have 
$t_{coll} \sim 8-26\; t_{ff}$ as Mach number varies from 1.05 to 7.
This time scale is much longer than the
observed values $\sim 2-5\; t_{ff}$ for prestellar cores. The reason for this 
disparity is that during the early part of its evolution, the central density of
the core is low, and it would not be identifiable within its surroundings. This is
evident in the low contrast between the center and the edge of the
core seen in the first
frames of Fig. \ref{fig:BE_shock_1.05} - \ref{fig:BE_shock_7.0}. Only when the
core approaches collapse does the center-to-edge contrast become large. For 
example, for the $M_a = 1.05$ model, the center-to-edge density ratio $\rho_c/\rho_{edge}$
reaches $2$ at $\tau = 0.78$ and reaches $5$ at $\tau=0.92$; this can be compared to the total
time until collapse, $\tau_{coll}= 1.027$ for this model. For the whole set of models,
the observable fraction of the pre-collapse core life is $30-50$\% if we choose 
$\rho_c/\rho_{edge} \geq 2$, or $10-20$\% if we choose $\rho_c/\rho_{edge} \geq 5$.
Taking the period when $\rho_c/\rho_{edge} \geq 2$ or $\rho_c/\rho_{edge} \geq 5$
as the period over which a core could be observable, e.g. in submilimeter continuum, 
our simulations give $t_{observable} \sim 2-13\, t_{ff}$ or 
$t_{observable} \sim 1-6\,t_{ff}$, respectively. The latter is consistent with
observed estimates.

Another observable aspect of prestellar cores is their density structure. As
discussed above, for each solution at the times shown in Fig. \ref{fig:BE_shock_1.05} 
- \ref{fig:BE_shock_7.0}, we fit a BE sphere profile.
The density in code units $\rho_0$ is 
$D(\xi)=\rho/\rho_0$ and the radius in code units is $\xi= r (4\pi G\rho_0)^{1/2}/a$.
When the central density is instead used to normalize, the density and radius
variables are:
\begin{equation}
\widetilde{D} = \rho/\rho_c = D\rho_0/\rho_c\, 
\end{equation}
and 
\begin{equation}
\widetilde{\xi}= r \frac{(4\pi G \rho_c)^{1/2}}{a} = \xi (\rho_c/\rho_0)^{1/2}\,.
\end{equation}
For a BE sphere with sound speed
$a_{BE}$, the density profile 
normalized by the central density is $\widetilde{D}_{BE}$, which is a function
of the scaled radius $\xi_{BE} = r (4\pi G\rho_c)^{1/2}/a_{BE} = \widetilde{\xi}\,a/a_{BE}$. To fit the density profiles in our simulations to a 
BE sphere, the only free parameter is the ratio of the isothermal
sound speed $a_{BE}$ of the BE sphere to the sound speed in the simulations, $a$. 
Thus, for any given density profile $\widetilde{D}$ we adjust the value of $a_{BE}/a$ 
until a good match between $\widetilde{D}_{BE}$ and $\widetilde{D}$ is obtained.
This procedure yields the fitted temperature ratio:
\begin{equation}
\frac{T_{BE}}{T_0} = \left(\frac{a_{BE}}{a}\right)^2\,.
\end{equation}
The values obtained for $T_{BE}/T_0$ are marked in Figures \ref{fig:BE_shock_1.05} - 
\ref{fig:BE_shock_7.0}. The range of values we find is 1.23 to 2.89.
This range is consistent with theoretical expectations. As the radius of the
BE sphere extends to infinity, the density profile approaches the
singular solution $\rho = 2 a_{BE}^2/(4\pi G r^2)$ \citep{shu77} corresponding
to $\widetilde{D}_{BE} = \rho/\rho_c = 2(a_{BE}/a)^2 \widetilde{\xi}^{-2}$.
The density profile of the collapsed core approaches the LP solution
$\widetilde{D}=8.86\widetilde{\xi}^{-2}$. To match the LP profile with the singular
profile therefore requires $2\,(a_{BE}/a)^2 = 8.86$, which corresponds to a 
temperature ratio $T_{BE}/T_0 = 8.86/2 =4.43$. Fits of core profiles to BE spheres
that indicate values of $T_{BE}$ up to 4.43 times the measured thermal temperature
therefore are expected if collapse has taken place; this does not by itself
indicate that magnetic support is present.

\subsection{Post-Collapse Evolution: Infall and Accretion Stages}

After the central density becomes singular, the evolution transitions to
the infall and accretion stages. For our simulations, we make this transition by implementing
an outflow boundary condition at the center when the central density reaches
$4\times 10^7 \rho_0$. The initial mass of the central protostar
is calculated by integrating the innermost part of the density profile where density
is between $[1\times10^7, 4\times10^7]\,\rho_0$.
The specific choice of this density does not significantly affect
$\tau_{coll},\, \xi_{coll}$, or the subsequent evolution since the central density 
increases dramatically only at the very end of the collapse stage.

At the beginning of the accretion phase, the material inside the shock
falls onto the protostar. The material approaching the protostar is in a
free-fall state \citep{hun77}.  The region of unsupported infall starts 
from the center and propagates outward, similar to the ``expansion
wave" described by \citet{shu77}. The density profile inside the
rarefaction wave changes from $\rho \propto r^{-2}$ to $r^{-3/2}$ and the
velocity profile changes from $v \propto r^0$ to $r^{-1/2}$. 
For our simulations, this infall stage ends as the infall rarefaction wave arrives 
at the shock front. This generally occurs very rapidly (in less than 10\% of
$\tau_{coll}$; see below).

For an initially static density profile $Ar^{-2}$, where $A = Ka^2/(4\pi G)$
and $K$ is a constant, if the gas pressure is negligible the average speed of the 
rarefaction wave is $(2\sqrt{2K}/\pi)\, a$.
For $K =8.86$, which is the LP profile, this yields $2.7a$. For the real
case, the initial velocity is nonzero and the gas pressure is
non negligible, so that the rarefaction wave propagates at a modified speed. 
For example, for the $M_a =1.05$ model, which has $\xi_{coll}= 0.29$ and
infall interval $\Delta \tau_{inf}= 0.092$, the average speed 
is $3.15a$. For $M_a =4.0$ and $7.0$, the measured average infall speed of the
infall rarefaction wave is $2.22a$ and $1.95a$ respectively.

After the infall rarefaction wave arrives at the shock front, 
the final accretion stage begins, with material initially 
outside the shock falling onto the protostar at supersonic
speeds. This process is similar to Bondi accretion, except that
the central mass is growing and the velocity field for our simulations is 
uniformly converging at large distance. During this stage,
the density and the velocity profiles vary $\propto r^{-3/2}$ and $r^{-1/2}$
respectively, corresponding to free-fall.

The typical density profiles and velocity profiles during the accretion 
stage for $M_a = 1.05$ and 4 models are 
shown in Figure \ref{fig:rho_v_acc}. Three different points in the evolution are
shown: the instant of core collapse, the instant when the infall rarefaction wave arrives
at the shock front, and a point during the late accretion stage. The transition 
from the LP profiles to the free fall profiles in density and velocity are clearly 
evident in the figure.

In a real system, the duration of the accretion stage depends on the environment of
the protostar, and how long the inflow that creates the core is maintained at large 
scales. To explore how the late-time evolution is affected by changes in the accretion,
we have conducted additional simulations in which the flow inward from the outer boundary
is halted at the end of the infall stage (i.e. when the rarefaction reaches the shock).
Suppression of inflow will affect the mass flux onto the protostar after
the rarefaction wave from the boundary reaches the central protostar.
We discuss comparison of these models to our standard models in section 4.4.

\subsection{Definition of Evolutionary Stages}

Based on the results presented in \S 4.1 and \S4.2, we have identified four main stages
of protostellar core formation and evolution in a supersonic turbulent
medium 
(see Figure \ref{fig:stages} for a schematic depiction):

1. {\it Core building ---}

Converging flows in a supersonic turbulent
medium collide, with post-shock compressed gas accumulating over time in stagnant, 
shock-bounded regions. If these dense regions are not destroyed by larger 
scale turbulence, the high 
density gas will undergo a long contraction process during which gas pressure 
competes with self-gravity. The typical character of this stage is that the velocity 
inside the dense gas is subsonic and increases linearly with distance from the center.
Since the center-to-edge density contrast is relatively low, these clumps are
gravitationally subcritical. Towards the end of this stage, when the center-to-edge
density contrast becomes appreciable, these objects would become observable as prestellar
cores. This core building stage lasts $>90$\% of $\tau_{coll}$; only
the last $10-20\%$ would be observable.

2.{\it Core collapse ---} 

The core built up during the first stage accumulates enough mass
that it becomes gravitationally supercritical, which we operationally define
as $r_{shock} > R_{BE,crit}$.
Self-gravity overwhelms the gas pressure, and the unstable core starts to collapse. 
The collapse begins near the shock front, where the imbalance between
gravity 
and
pressure gradient forces is greatest, and propagates inward. This collapse is
an ``outside-in'' process. During core collapse, the central density increases
dramatically and the inflow velocity becomes supersonic. As the collapse propagates 
inward, a density profile $\rho \propto r^{-2}$ is left behind and the velocity increases
toward the center. The end of this stage is defined by the instant of protostar formation
$\tau_{coll}$, when the wave of collapse has reached the center of the core.
The density profile approaches the LP profile, $\rho = 8.86 a^2/(4 \pi G r^2)$.
The velocity in the interior of the collapsed core approaches a uniform value 
comparable to that in the LP solution $-3.28a$. The central density is high enough 
to be optically thick and a protostar forms. Because of their large central volume
densities, prestellar cores during this collapse stage would be observed as having 
high peak submm flux densities. This stage lasts for a time 
$\Delta \tau_{supcrit}$, less than 10\% of the prestellar core lifetime $\tau_{coll}$.

3. {\it Envelope infall ---} 

During this stage, the high density material inside 
the shock front falls to the protostar
(or, if angular momentum were included, a circumstellar
disk). This stage starts at
the instant of the protostar formation and ends at the instant when
the gravitational rarefaction wave reaches the shock front, clearing
out the remnants of the dense gas that accumulated during core building. During
this stage, the density and velocity profiles in the interior change from
LP profiles to free-fall profiles. Since the core contains an embedded protostar,
the system would be observationally classified as Class 0/I. This stage lasts
for a time $\Delta t_{inf}$
less than 10\% of the prestellar core lifetime.
In a real system, the fraction of the envelope mass that eventually reaches
the center would depend on the details of the protostellar wind, which
would sweep up at least the polar portion of the envelope, creating an
outflow.

4. {\it Late accretion ---} 

During this stage, material from the ambient environment
directly accretes to the protostar (or, more realistically, a disk). 
Accretion during this stage is similar to Bondi accretion, with free-falling density 
and velocity profiles $\rho \propto r^{-3/2}$ and $v \propto r^{-1/2}$ over the whole core 
region. For a real system, the duration of
the late accretion stage, and hence the final stellar mass, depends on ambient 
conditions far from the protostar.  In a real system, the potential of
this stage to contribute significantly to the stellar mass 
would also depend on protostellar winds, which can reverse the accretion.

Although we have identified these stages based on idealized spherically symmetric 
models with constant gaseous inflow rates at large distance, we expect that the same
stages would be present, in modified form, under more realistic conditions. Based
on our simulations, the ratios $\Delta \tau_{supcrit}/\tau_{coll}$ and
$\Delta \tau_{inf}/\tau_{coll}$ decrease with increasing $M_a$, as shown in Figure 
\ref{fig:dtau}. The supercritical stage and the infall stage have similar durations,
and range from  $9-3\%$ of the prestellar core lifetime $t_{coll}$, which itself 
ranges from $8-26\,t_{ff}$. The supercritical collapse period (stage 2) 
$\Delta t_{supcrit}$ and the infall period (stage 3) $\Delta t_{inf}$ thus both range over
$\sim 0.8-1\,t_{ff}$, as shown in Figure \ref{fig:dtau}. While the time to reach collapse
would differ for nonspherical or nonsteady converging large-scale flows, we expect
that the character of the evolution would not. We also expect that the ratios 
$\Delta t_{supcrit}/t_{ff}$ and $\Delta t_{inf}/t_{ff}$ would remain
order-unity.

\subsection{Evolution of Mass Accretion Rates}

\begin{figure}[ht]
\centerline{
\includegraphics[width=0.8\textwidth]{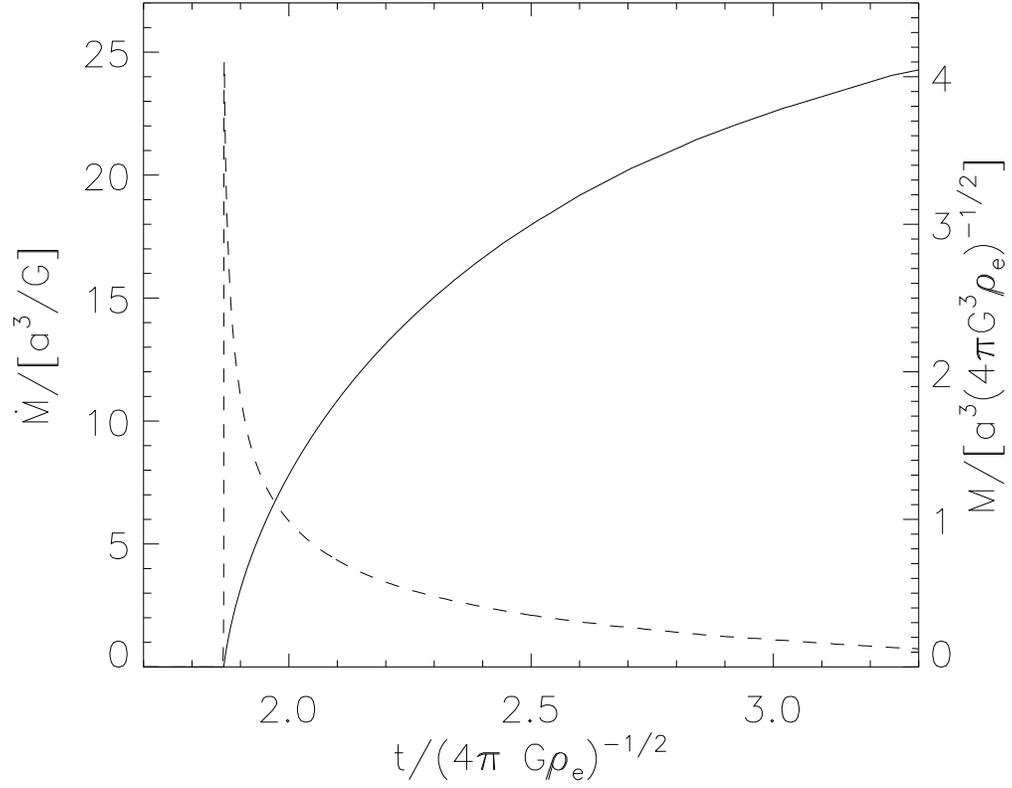}
}
\caption{{\it Dashed curve, left scale:} 
      The temporal evolution of the mass accretion rate at 
      inner edge of the grid, for collapse initiated from a critical BE
      sphere (see eq. \ref{accr_def} for units). 
      {\it Solid curve, right scale:} Evolution of the 
      central point mass (see eq. \ref{M0_def} for units). 
      Time is shown in units scaled by the density at the outer edge (see eq. \ref{t0_def}).}  
\label{fig:mfl_pre}
\end{figure}

\begin{figure}[ht] 
\centerline{
\includegraphics[width=1.0\textwidth]{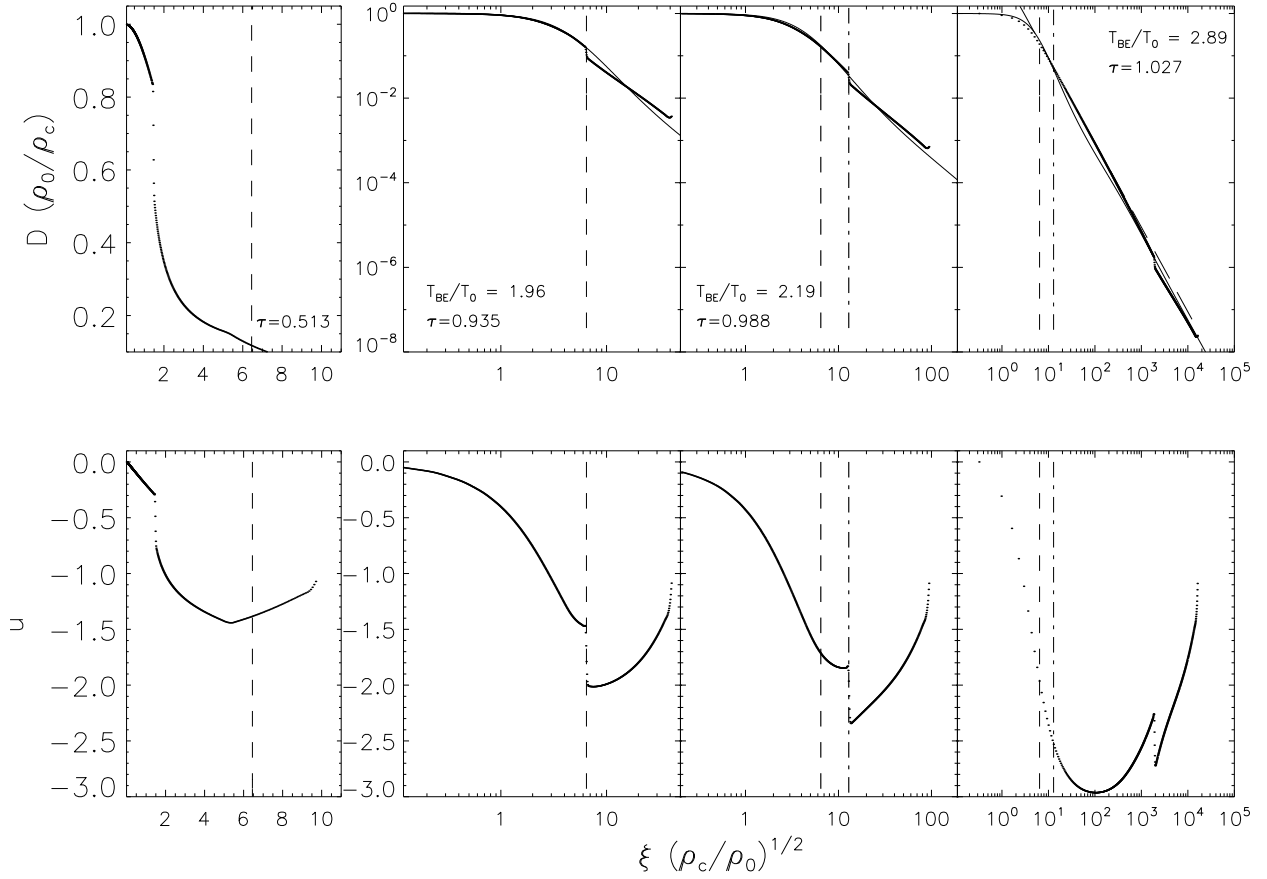}
}
\caption{Density and velocity profiles ({\it dotted curves}) for
converging-flow model with $M_a = 1.05$, at times $\tau$ as noted in 
the upper panel for each pair. The radius is normalized by
the central density, i.e. the abscissa is $r(4 \pi G \rho_c)^{1/2}/a$.
For the three upper panels on the right, the solid line is a fit to a BE
sphere with temperature $T_{BE}$, with the fitted temperature noted in
each panel. Dashed vertical lines denote the critical radius of a
BE sphere with the same central density and sound speed $a$. 
Dot-dashed vertical lines
mark twice this critical radius.
The time for the leftmost pair is half of the collapse time $\tau_{coll}$. The
time for the second and third pairs are when the shock reaches the
critical BE radius and twice that value. The time for the last 
pair is the instant of collapse $\tau_{coll}$ (defined in the simulations as 
$\rho_c/\rho_0=4\times 10^7$). The top-right panel shows with a 
dashed diagonal line the LP density profile $D=8.86 \xi^{-2}$.
}
\label{fig:BE_shock_1.05}
\end{figure}

\begin{figure}[ht]
\centerline{
\includegraphics[width=1.0\textwidth]{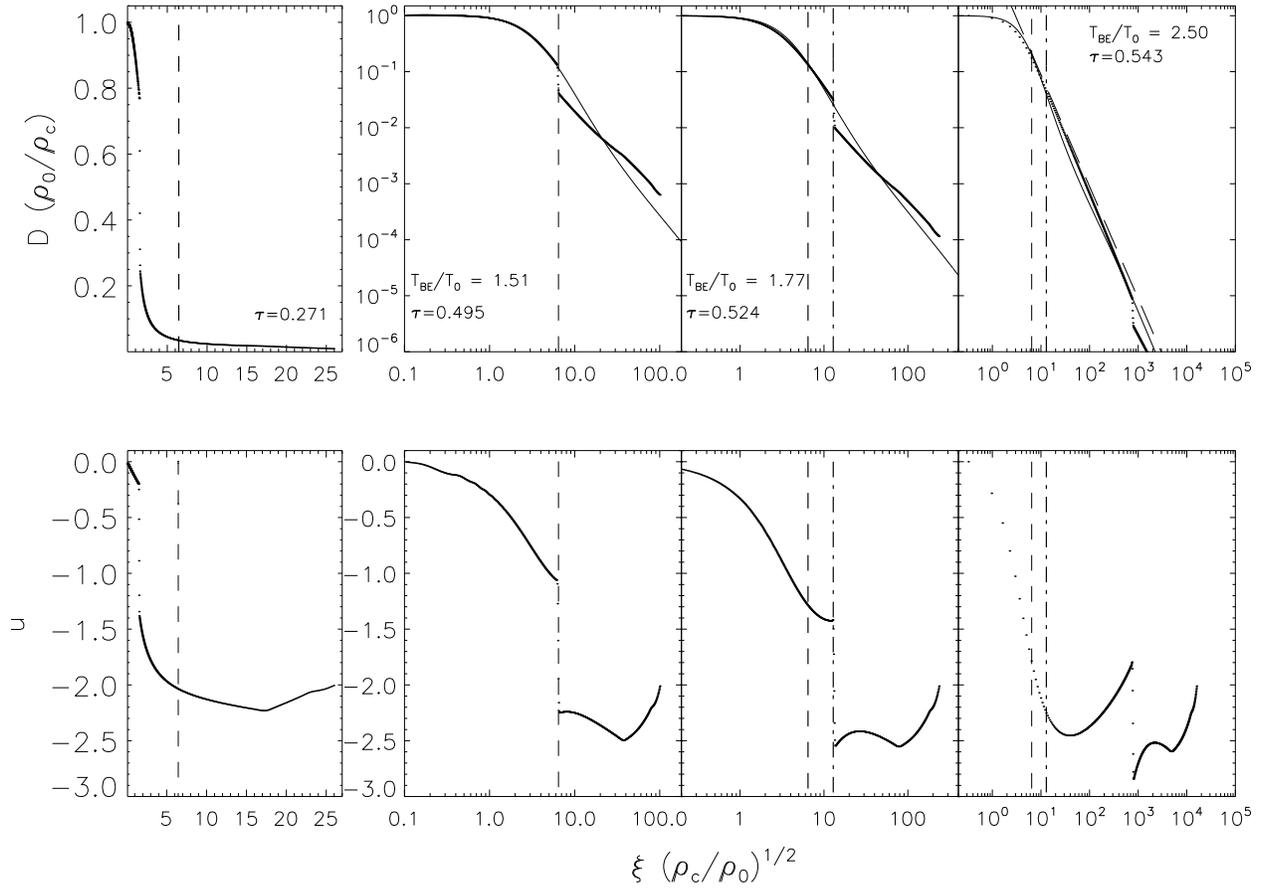}
}
\caption{Same as Figure \ref{fig:BE_shock_1.05}, for inflow Mach number $M_a=2.0$.}
\label{fig:BE_shock_2.0}
\end{figure}

Figure \ref{fig:mfl_mass} shows the temporal evolution of
the mass accretion rate and the total integrated mass of the central protostar
for $M_a = 1.05, 2, 4$ and $7$. These can be compared to the mass accretion 
rate and the integrated central mass for an initially static critical
BE sphere, as shown in Figure \ref{fig:mfl_pre} (see also \citet{vor05}, who
show similar accretion histories to Figure \ref{fig:mfl_pre}). For both the initially-static
collapse and our models that allow for core building from supersonic flows, there
is a sharp early peak in the accretion rate. The rise starting from $\dot{M}=0$ 
corresponds to the moment of protostar formation at $\tau_{coll}$. 
The smooth decline that follows
(ending at the points marked ``i'' in Figure \ref{fig:mfl_mass}) corresponds to
the infall stage, as the gravitational rarefaction wave propagates outward. 
At late times, however,
the accretion differs for the initially-static vs. dynamically-built cores. 
For an initially-static unstable BE core, the late accretion steadily declines over
time (Fig. \ref{fig:mfl_pre}). In contrast, for cores formed in the $M_a = 1.05$
model (i.e. barely supersonic inflow), the late time accretion is nearly constant,
and for cores formed in large $M_a$ models, the late accretion rate increases
over time (thin lines in Fig. \ref{fig:mfl_mass}).

The early-time peak accretion rates can be compared to the predictions of 
self-similar models. For the LP profiles at the instant of core collapse, 
$D=8.86\xi^{-2}$ and $u = -3.28$ give $\dot{M}=29.1 a^3/G$, while the self-similar 
solution for the accretion phase in \citet{hun77} predicts 
$\dot{M}= 46.195 a^3/G$. In fact, we do see a jump in $\dot{M}$ above $29.1 a^3/G$
as the evolution transits from the collapse stage to the infall
stage. This phenomenon is most clearly evident for the $M_a = 1.05$ model, which
has the highest resolution of the central region because the shock strength is lower
than in the high $M_a$ models, yielding a larger core (see Fig. \ref{fig:rho_v_collapse})
at the instant of collapse.

The detailed behavior of $\dot{M}$ during the late accretion stage can be 
understood in terms of various transitions that occur.
For $M_a =1.05$, the accretion rate (see in Fig. \ref{fig:mfl_mass}) starts to increase 
from point $i$ until to point $p$, and then decreases.
The increases from $i$ to $p$ occurs as gas stored between the shock front and the outer
boundary collapses into the center. The point $p$ represents the instant when the
density in the whole outer region reaches a profile $\rho \propto r^{-3/2}$. After
point $p$, the gravitational rarefaction has reached the boundary, and subsequent
accretion is limited by the inflow rate imposed at the outer boundary.
For the $M_a =2, 4$ and $7$ models shown in Figure \ref{fig:mfl_mass}, the mass accretion
process between point $i$ and $p$ is similar to that of $M_a = 1.05$.
However, there is additional transitory behavior before the accretion rate decreases to the
inflow rate imposed at the boundary. At the instant corresponding to point $p$, 
the rarefaction has produced  $\rho \propto r^{-3/2}$ over the whole region. 
But the velocity profile has $v \propto r^{-1/2}$ only over the inner region. 
During the stage between point $p$ and $q$, 
the density profile stays almost unchanged but the velocity profile evolves to reach 
$v \propto r^{-1/2}$ everywhere (see also Fig. \ref{fig:rho_v_acc}).
After point $q$, the accretion rate decreases to the imposed inflow rate.
The stage between $p$ and $q$ is most obvious for the $M_a =4.0, 7.0$ models.

As mentioned above, we have also performed models in which inflow to the grid is
halted after the point when the rarefaction reaches the shock. The resulting late-stage
accretion (see Fig. \ref{fig:mfl_mass}) is initially the same as in our standard models,
but then declines over time, after the rarefaction wave reaches the boundary.
  
The mass of the protostar at the end of the infall
stage, and the total core mass $\mathrm{M}_{core}$ inside the shock at instant of core
collapse, are shown in Figure \ref{fig:core_mass}. For comparison, 
we also show the critical BE sphere mass (see eq. \ref{massBE}) using the post-shock
density at the time of collapse for $\rho_{edge}$; these are much lower than the actual 
core mass. Because there is continued mass passing through the shock during the infall
stage, the post-infall protostellar mass is slightly higher than the core mass inside
the shock at the time
when core collapses. As the Mach number increases, the post-infall protostellar mass 
and the core mass at $\tau_{coll}$ both decrease.
The protostellar mass ranges over $0.06 - 8.8\Msun$ and the core mass at $\tau_{coll}$ 
ranges over $0.05 - 7.5\Msun$, taking $n_H = 100 \pcc$ for the ambient density.

\section{Summary and Conclusions}
Star formation takes place in GMCs pervaded by supersonic turbulence, and
theoretical models of prestellar (and protostellar) cores must take
these large-scale supersonic flows into account.  Here, we have
developed models in an idealized, spherically-symmetric framework that
nevertheless captures key aspects of the real situation, enabling us
to identify and analyze the main stages of core formation and
evolution in a dynamic environment.

Our models differ from previous studies of core evolution in that the
cores are not present as either stable or unstable density
concentrations in our initial conditions -- the initial density is
everywhere uniform. Instead, during the first evolutionary stage cores
are built ``from scratch'' by the collision of converging supersonic
flows.  The boundaries defining the outer edge of a core -- where the
density drops -- correspond to a shock front across which the
temperature is constant and the mass flux is nonzero.
The shock front propagates outward, with the mass of the
post-shock dense region growing in time.  Initially, the core is
essentially uniform.  Over time, the mass grows sufficiently so that
the core becomes centrally stratified due to
self-gravity. Observationally, the latter part of this ``core
building'' stage corresponds to prestellar cores that have low to
intermediate peak brightness.  The period over which
$\rho_c/\rho_{edge}\ge 5$, and a core would be clearly identifiable in
observations, amounts to 1 -- 6 $t_{ff}$, with the free-fall time
defined using the mean core density.

When the center-to-edge density contrast exceeds $\sim 10$, the core
becomes supercritical and a stage of violent ``outside-in'' collapse
ensues.  The density profile throughout the core approaches $\rho
\propto r^{-2}$, and a protostar forms at the center.  We define the
instant that collapse is complete and a protostar forms as $t_{coll}$,
or $\tau_{coll}$ in our dimensionless variables.  Although the
central density becomes very large, the wave of outside-in collapse
still leaves most of the core mass in the outer parts.
Observationally, this core collapse stage corresponds to prestellar 
cores that have high peak brightness.  The period $\Delta
t_{supcrit}$ over which cores are supercritical, undergoing
outside-in collapse, amounts to less than 10\% of $t_{coll}$, or 0.8
-- 1 $t_{ff}$.

The third stage of evolution is governed by an ``inside-out'' wave of
gravitationally-driven rarefaction propagating from the center of
the core to the shock front that defines the core's outer edge.  The
accretion rate onto the star during this infall stage is initially
very high, but declines over time.  At the end of this infall stage,
the dense envelope built during the first stage has plunged into
the star.  The velocity and density profiles approach free-fall,
$v\propto r^{-1/2}$ and $\rho \propto r^{-3/2}$, respectively.
Observationally, this stage corresponds to Class 0/I embedded
protostars.  The period $\Delta t_{inf}$ over which cores undergo
this inside-out infall is similar to the duration of the previous
stage, $\Delta t_{supcrit}$, and comparable to $t_{ff}$.

During the final stage of evolution, there is no longer a massive
envelope.  The protostar can continue to accrete from the more
distant, lower-density gas in its surroundings.  The late-stage
accretion rate and total mass accumulated by the system 
depend on the large-scale environment,
rather than the properties established in the core during the building
stage.  Observationally, this stage corresponds to a non-embedded YSO
that may still be accreting from a disk.

Based on our simulations, our chief conclusions are as follows:

1. The initiation of star formation via outside-in core collapse, followed by
   inside-out envelope infall, appears to be very robust.  The dynamical
   behavior during these stages of evolution is very similar whether
   the core is initiated as an unstable equilibrium (as in previous
   models) or is built up dynamically through a shocked converging
   flow (as in the present work).  The Larson-Penston singular
   solution with $\rho =8.86 a^2/(4\pi G r^2)$ appears to be an ``attractor,''
   in that models initiated from stationary equilibria or with
   different supersonic converging velocities all arrive at this
   configuration at the moment of protostar formation.

2. Prior to the point at which cores become supercritical and outside-in
   collapse begins, the velocities interior to cores are subsonic,
   even if they are created by highly supersonic flows.  In fact,
   {\it higher} inflow velocities produce {\it lower} post-shock
   velocities within the dense core (see Figs. 2-5 and eq. 23).  This
   result is consistent with observations showing that dense cores are 
   quiescent in their interiors (see e.g., \citealt{1983ApJ...270..105M,
   1998ApJ...504..223G,
  2001ApJS..136..703L, 2002ApJ...572..238C,
   2007ApJ...668.1042K,
   2007A&A...472..519A,2008ApJ...672..410L}).

3. Throughout both the core-building and core-collapse stage, density
   profiles for cores formed by shocked converging flows can be fit by
   Bonnor-Ebert profiles, but with fitted temperatures $T_{BE}$ larger
   than the true temperature $T$.  The range of temperatures fitted for our
   models with Mach numbers up to 7 is $T_{BE}/T=1.2-2.9$.  This
   result is consistent with observational findings (\citealt{kan05, kir05}) 
   that effective
   temperatures greater than directly-measured values (from fitting
   SEDs) are usually required in order to fit BE spheres to observed
   prestellar cores.  The largest possible ratio that could be
   obtained for an isothermal spherical flow is $T_{BE}/T=4.43$, so
   that any observed ratio larger than this suggests that magnetic
   fields contribute appreciable support, or else the core is anisotropic.
   \citet{dap09} have also recently pointed out that the temperature fit 
   based on matching a BE profile may be significantly higher
   than the true kinetic temperature, for clouds in collapsing stages.

4. At the time of collapse, for all Mach numbers the core size and
   mean density are closely related.  We find that 
   $R_{coll} \approx 4 a (4 \pi G \rho_{mean})^{-1/2}$ within 15\% for Mach
   numbers $M_a=1.05 - 7$.  This radius is $\sim 50\%$ larger than the
   critical radius of a BE sphere with the same mean internal density.

5. As $M_a$ increases, and assuming a given ambient medium density
   $\rho_0$, the time to reach collapse $t_{coll}$ is shorter, the
   physical size of the core at $t_{coll}$ is smaller, the mean
   internal density at $t_{coll}$ is higher, and the mass of the core
   at $t_{coll}$ is lower.  For high Mach number, the collapse time
   and collapse radius are related by $t_{coll} \approx 2.6 R_{coll}
   M_a/a$.  The range of core masses at the time of collapse at
   different $M_a$ is consistent with observed core masses, although
   the specific dependence of core mass on $M_a$ found in the present
   work may be sensitive to the spherical converging-flow geometry we
   have adopted.

6. The durations of the infall stage and collapse (supercritical)
   stage of evolution are comparable for all $M_a$, and are close to
   $t_{ff}$.  This is consistent with observations indicating similar
   lifetimes for prestellar cores and embedded Class 0/I accreting
   protostars \citep{ 1986ApJ...307..337B,
   1999ApJS..123..233L, 2000MNRAS.311...63J,kir05,2007ApJ...656..293J,
   2007A&A...468.1009H, 2008arXiv0805.1075E,2008arXiv0811.1059E}. 
   These stages are preceded by an extended
   core-building stage, during most of which the core would not be
   observable because its center-to-edge density contrast is low.

7. The mass accretion rate onto the protostar (or, more realistically,
   star-disk system if angular momentum were included) peaks at the
   beginning of the infall stage at a value $\gg a^3/G$, and then
   declines steeply afterwards as the material stored in the envelope
   is exhausted.  This result appears to hold regardless of how cores
   form, as it is consistent with earlier work (see e.g., \citealt{fos93,ogi99,
   hen03,mot03,vor05}) for cores
   initiated from unstable equilibria or which undergo 
   externally-induced compression.  Later accretion from the
   ambient medium depends on how long the large-scale cloud maintains
   a focused converging flow.

As noted above, some of our specific conclusions are likely to change
for non-spherical geometry, and for time-dependent rather than steady
large-scale inflow.  Furthermore, other elements that are present in
real star formation have been entirely omitted in these models;
these elements include rotation, which would lead to disk formation;
protostellar winds, which would sweep up and remove a portion of the
envelope during the infall stage, and could prevent late accretion
altogether; and magnetic fields, which would alter the timescales and
details of the evolutionary stages.

We expect, however, that many of our basic
results will carry over even if the idealizations we have adopted are
relaxed.  While large-scale supersonic converging flows in real GMCs
are not generally spherical, the association of observed cores with
high density surroundings suggests that the dense gas in post-shock
stagnation regions is still the raw material out of which cores are
built.  We expect that in general core masses and collapse timescales will
decrease with increasing density of the post-shock flow, which itself
increases with increasing Mach number. Preliminary simulations of 
planar converging flows that we have conducted 
indeed bear out this expectation, showing
$M_{core} \propto M_a^{-1}$.  For planar converging flows, many cores
simultaneously grow and then collapse in the post-shock gas layer; unless 
this sheet were viewed exactly edge-on, the density jump 
at the shock front would not be apparent, and cores would be
seen as surrounded by moderate-density gas.

We also
expect outside-in collapse followed by inside-out infall to be a
generic feature of core evolution. Although the duration of this
pressure-mediated collapse is $\sim t_{ff}$, it is {\it unlike}
free-fall collapse in a crucial way: the core does not remain
nearly-uniform.  We speculate that the development of stratification
during inside-out collapse will suppress growth of perturbations and
sub-fragmentation for the non-spherical case.  Even though the
increase in density implies that the local Jeans mass becomes smaller
and smaller, this is only true in the very center of the core.
Instead, we expect that collapse of cores built within shocked converging
flows will produce single systems, which may be binary (or multiple)
if the angular momentum is sufficient.  Non-spherical converging flows
that create sheets and filaments of shocked gas and produce many such
cores simultaneously could be the progenitors of stellar clusters.

\acknowledgements
This research was supported by grant NNG-05GG43G from NASA.  We are
grateful to Phil Myers, Shantanu Basu, and the referee for careful
reading and helpful comments on the manuscript.

\newpage

\clearpage

\begin{figure}[ht] 
\centerline{
\includegraphics[width=1.0\textwidth]{fig04.epsi}
}
\caption{Same as Figure \ref{fig:BE_shock_1.05}, for inflow Mach number $M_a=4.0$.}
\label{fig:BE_shock_4.0}
\end{figure}
\begin{figure}[ht]  
\centerline{
\includegraphics[width=1.0\textwidth]{fig05.epsi}
}
\caption{Same as Figure \ref{fig:BE_shock_1.05}, for inflow Mach number $M_a=7.0$.}       
\label{fig:BE_shock_7.0}
\end{figure}

\begin{figure}[ht]  
\centerline{ \includegraphics[width=0.8\textwidth]{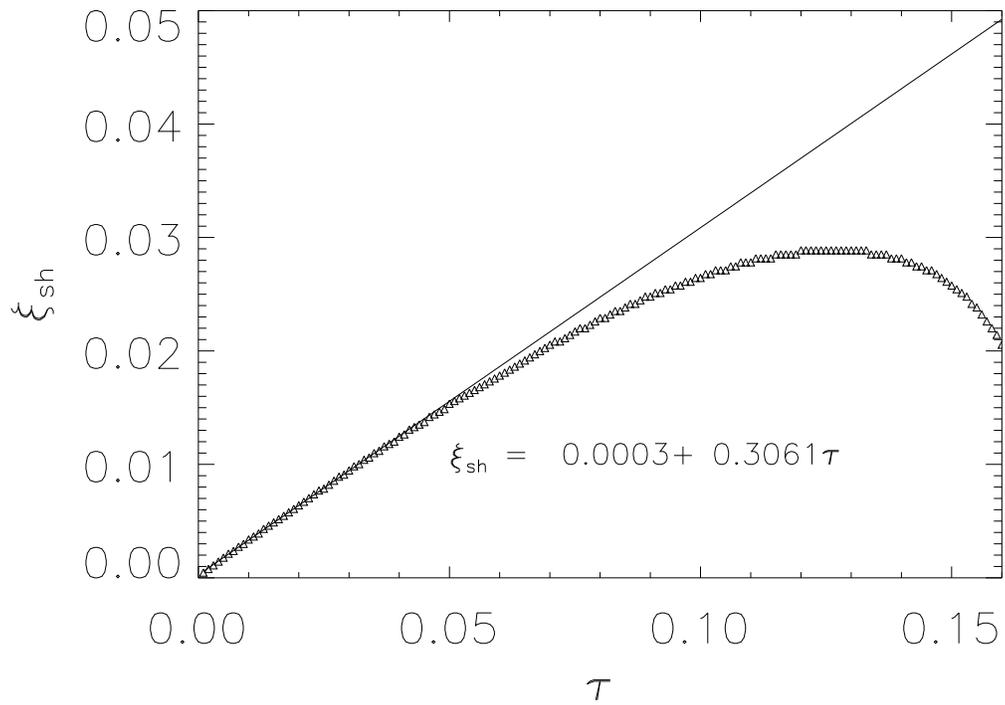}}
\caption{The shock front position versus time for $M_a =4$. The solid 
line is the best fit to the linear part, i.e. a constant-speed outward-
propagating shock at early times. Collapse occurs for this model at
$\tau_{coll}=0.16$.}
\label{fig:shkfrps}
\end{figure}

\begin{figure}[ht]  
\centerline{
\includegraphics[width=0.8\textwidth]{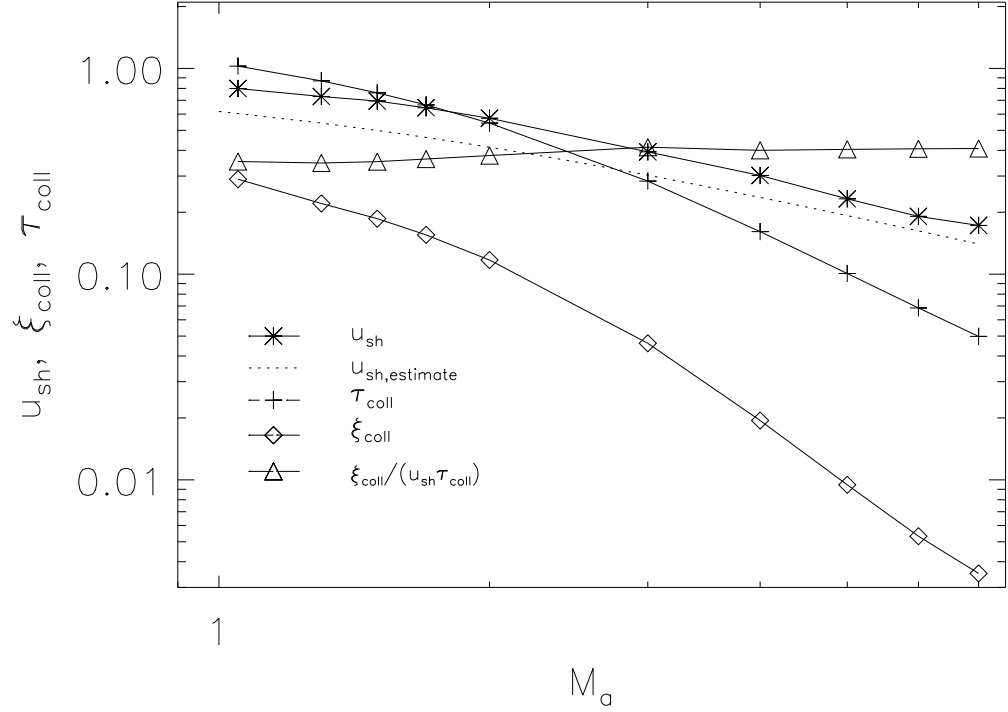}
}
\caption{The speed of shock front at early times (asterisks), the radius of the core 
at the instant of collapse (diamonds), and the time at which collapse 
occurs (plus signs) as a function of $M_a$. Triangles show the ratio 
of $\xi_{coll}/(u_{sh}\tau{_{coll}})$, which is nearly constant, ranging
from 0.34 to 0.42. The dotted line is the analytic estimate for $u_{sh}$
given in equation (\ref{eq:ush}).}
\label{fig:u_xi_tau_ori}
\end{figure}

\begin{figure}[ht]  
\centerline{
\includegraphics[width=0.8\textwidth]{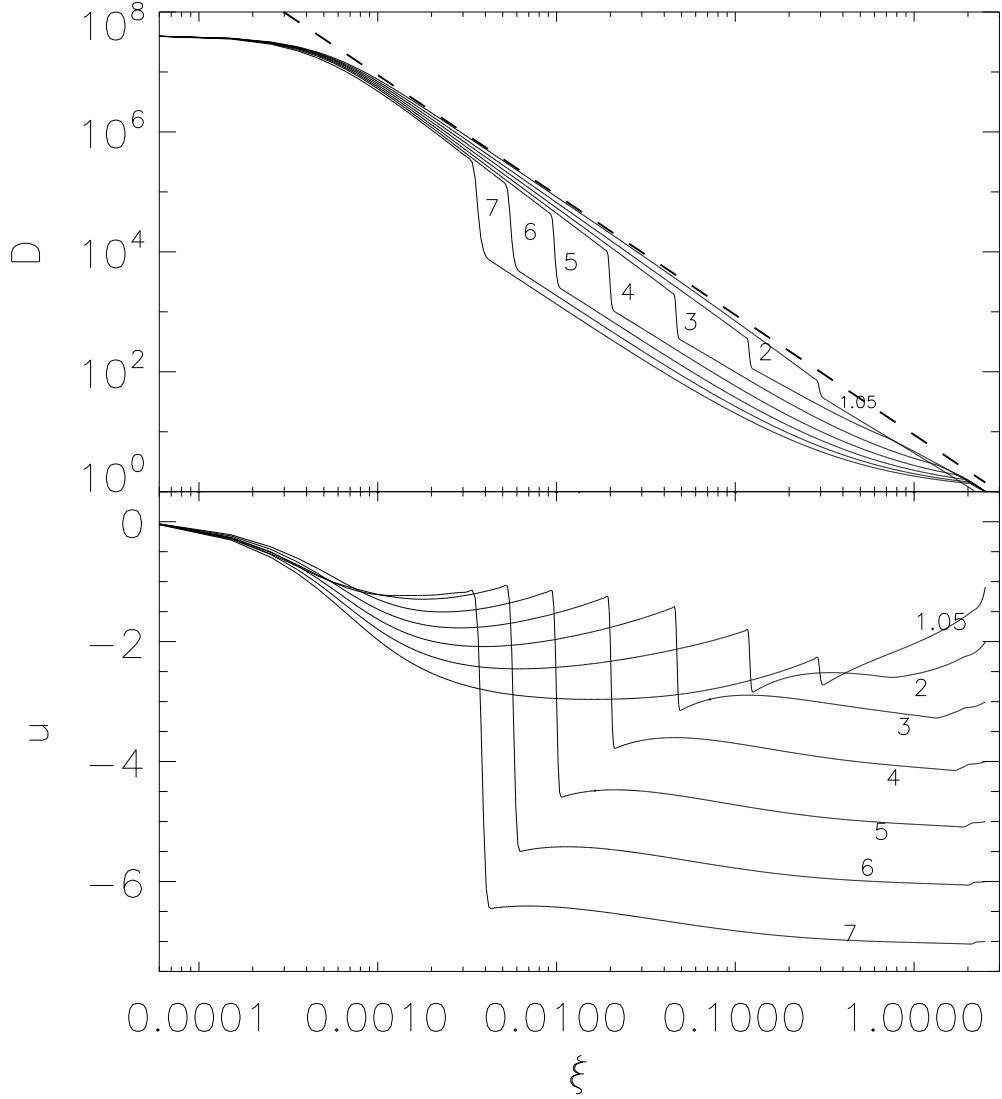}
}
\caption{The density profiles (top panel) and the velocity profiles (bottom panel)
at the instant of collapse for simulations with different Mach number $M_a$ (as
labeled). 
The thick dashed line in the top panel is the LP density profile $D=8.86\xi^{-2}$.  
For higher $M_a$ cases, the shock front is at smaller radius (in units of $r_0$) 
and the post-shock speeds in the dense core are lower in magnitude. Units of 
length and velocity are given by equations (\ref{r0_def}) and (\ref{isosp_def}),
respectively. Density is in units of the GMC ambient value.} 
\label{fig:rho_v_collapse}
\end{figure}

\begin{figure}[ht]  
\centerline{
\includegraphics[width=0.8\textwidth]{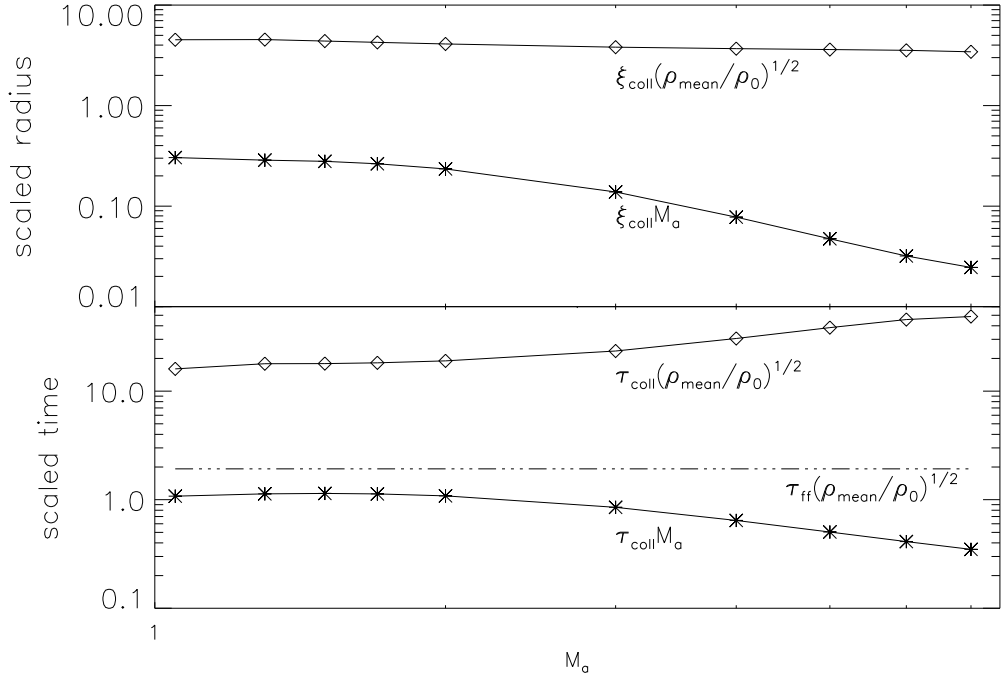}
}
\caption{The renormalized radius (top panel) and collapse time (bottom panel) 
of cores versus $M_a$. Diamonds show quantities normalized using the mean core density
and asterisks show quantities normalized using the Mach number (see text).
The dot-dashed line in the lower panel shows the free-fall time at density $\rho_{mean}$       
in units $(4\pi G \rho_{mean})^{1/2}$, i.e. $\tau_{ff}(\rho_{mean}/\rho_0)^{1/2}
= \pi (3/8)^{1/2}$.}
\label{fig:u_xi_tau_nor}
\end{figure}

\begin{figure}[ht] 
\centerline{
\includegraphics[width=1.0\textwidth]{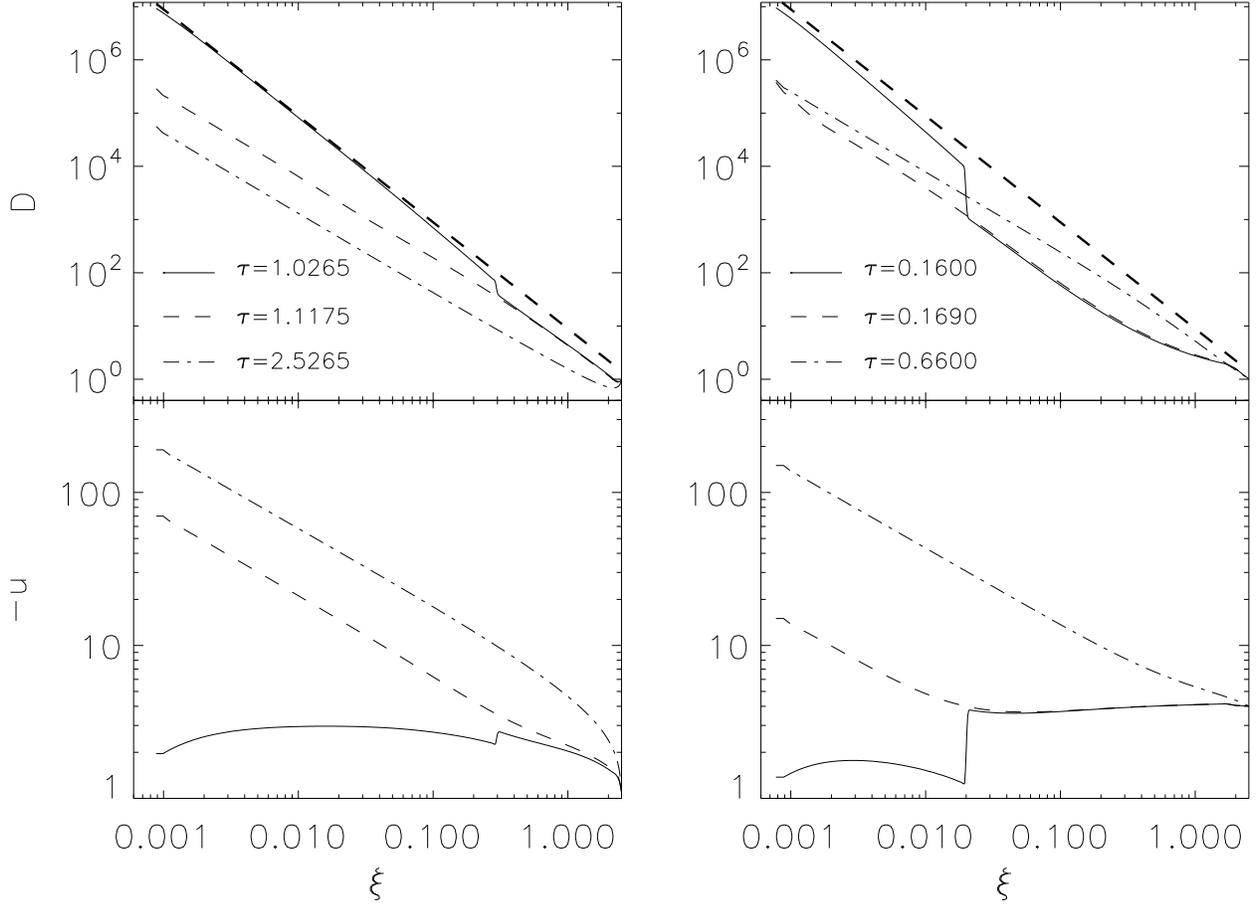}
}
\caption{The density and velocity profiles during
the accretion phase for $M_a = 1.05$ (left) and $M_a= 4$ (right) models. 
Density profiles (top) and the velocity profiles (bottom) 
are each shown at three different instants: solid lines show the instant of 
core collapse, dashed lines show the instant when the gravitational
rarefaction wave arrives at the shock front, and dot-dashed lines
show the profiles at a late accretion stage. Numbers in the figure
show the corresponding time for each instant. In the upper panels, the LP
density profile $D=8.86\xi^{-2}$ is plotted for reference with thick 
dashed lines. The transitions from $D\propto \xi^{-2}$ (early) to 
$D\propto \xi^{-3/2}$ (late) and $u\propto \xi^0$ (early) to 
$u\propto \xi^{-1/2}$ (late) are evident.}
\label{fig:rho_v_acc}
\end{figure}

\begin{figure}[ht] 
\centerline{
\includegraphics[width=1.0\textwidth]{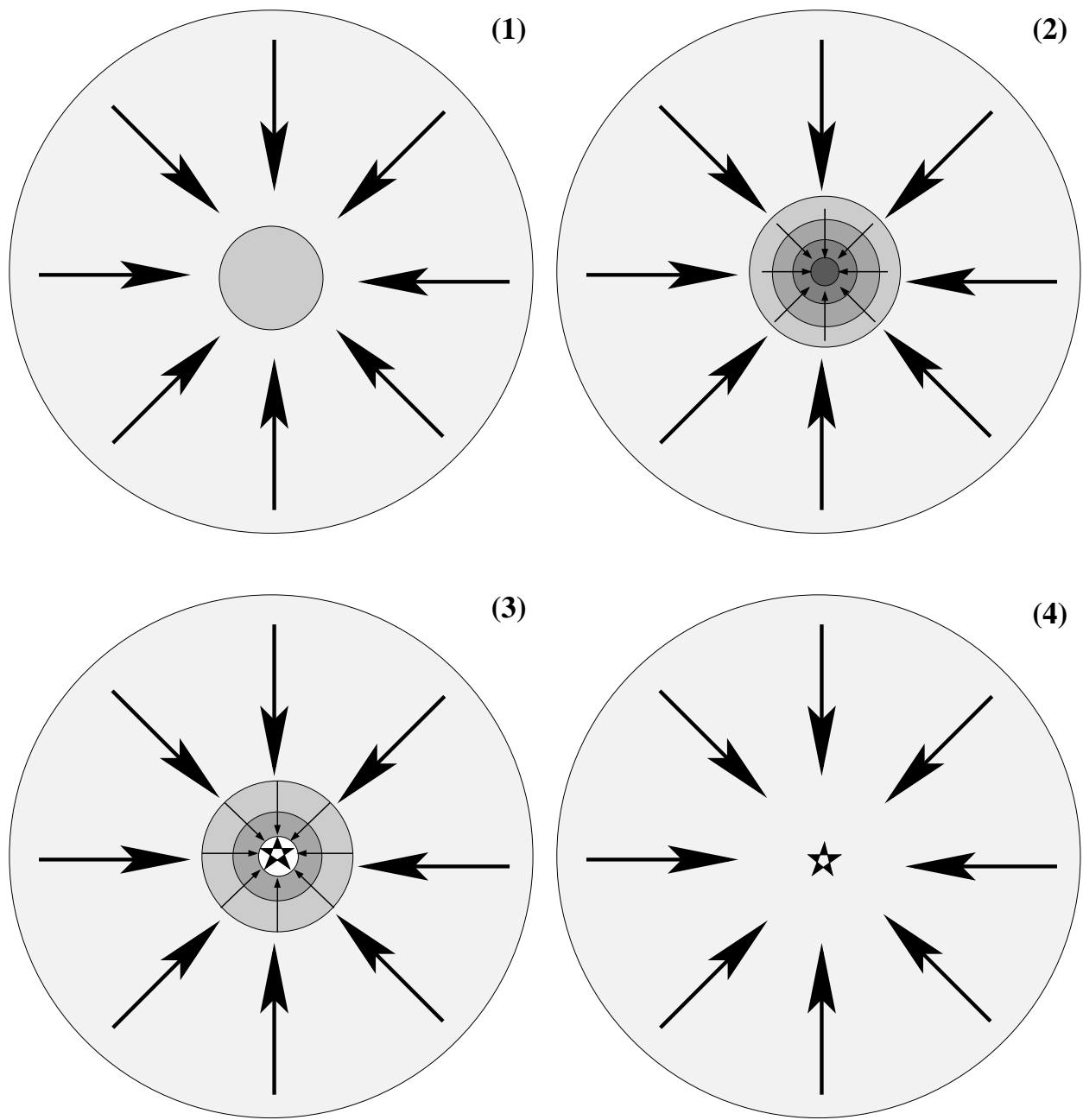}
}
\caption{ The four stages of core evolution in the idealized models of
this paper. (1) Core building via supersonic converging flow, yielding
a stagnant, shock-bounded dense region.  (2) Core collapse,
propagating from outside to inside, leading to a density profile $\rho
\propto r^{-2}$. (3) Envelope infall, propagating from inside to
outside, and resulting in free-fall onto the protostar.  (4) Late
accretion of ambient gas onto the protostar.  
For real systems,
evolution would be modified in several ways: converging flows would be
non-spherical, angular momentum would lead to disk formation, and
outflows would contribute to clearing the envelope.}
\label{fig:stages}
\end{figure}

\begin{figure}[ht]  
\centerline{
\includegraphics[width=0.8\textwidth]{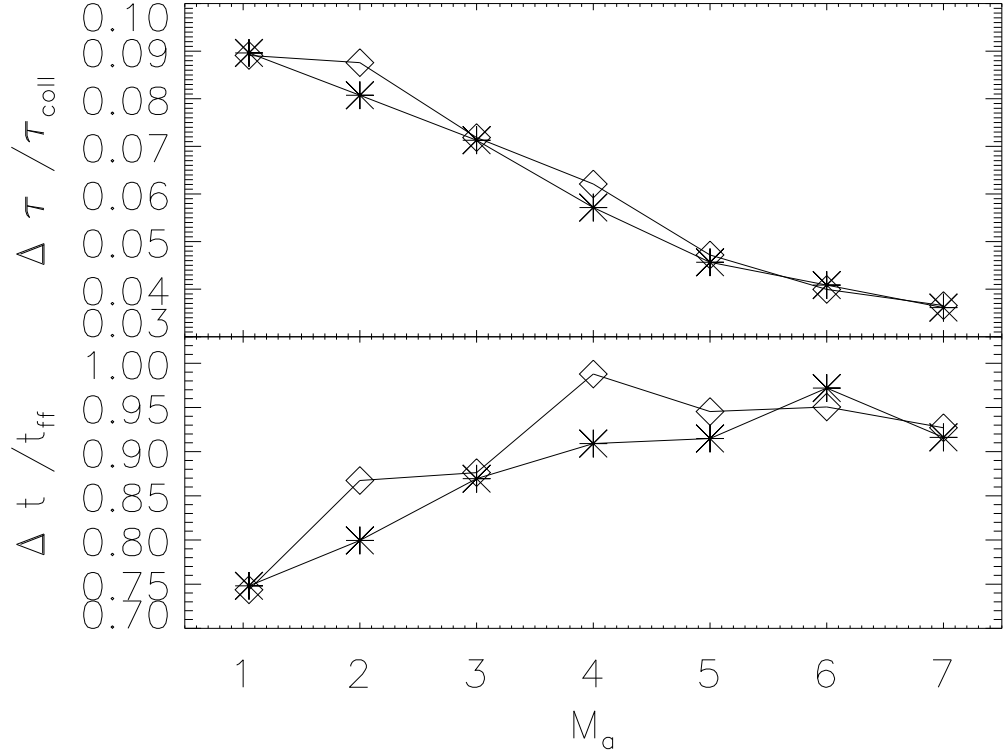}
}
\caption{Top panel shows the ratio of the duration of the supercritical 
collapsing stage $\Delta \tau_{supcrit}$ (asterisks) and the post-collapse infall 
stage $\Delta \tau_{inf}$ (diamonds) 
to the prestellar lifetime $\tau_{coll}$ of cores, as a function of Mach number. 
Bottom panel shows ratios $\Delta t_{supcrit}/t_{ff}$ (diamonds) and 
$\Delta t_{inf}/t_{ff}$ (asterisks), which range from [0.8, 1], as a function 
of Mach number; here $t_{ff}$ is computed using the mean density inside the shock
at $\tau_{coll}$.}
\label{fig:dtau}
\end{figure}

\begin{figure}[ht]  
\centerline{
\includegraphics[angle=270,width=0.8\textwidth]{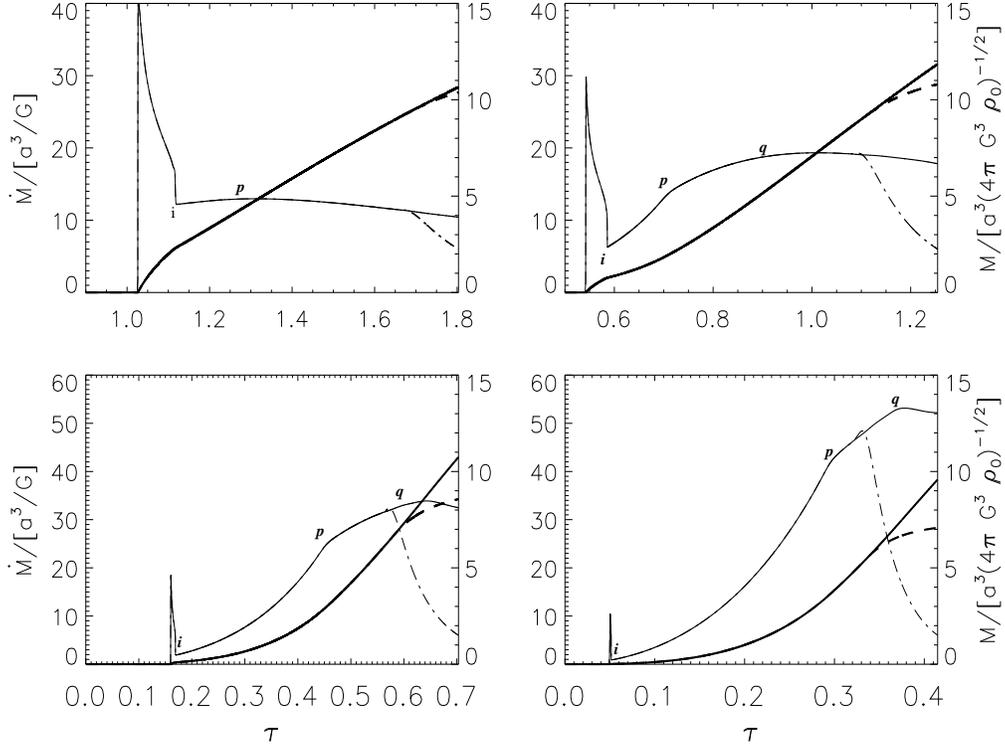}
}
\caption{The temporal evolution of the mass accretion rate (thin solid lines;
left axis) and the integrated mass of protostar (thick solid line; right axis) 
for models with Mach number $M_a = 1.05, 2, 4$ and $7$, as labeled. Corresponding dashed lines
show the results for models in which the inflow to the grid is suppressed after the
end of the infall stage. The point marked $i$ represents the end of
the infall stage, $p$ represents the instant when the density profile 
reaches $\rho \propto r^{-3/2}$ everywhere, $q$ represents the instant
when the velocity profile reaches $v\propto r^{-1/2}$ everywhere. Units of time,
mass, and accretion rate are given by equations (\ref{t0_def}), (\ref{M0_def}) and
(\ref{accr_def}), respectively.}
\label{fig:mfl_mass}
\end{figure}

\begin{figure}[ht]
\centerline{
\includegraphics[width=0.8\textwidth]{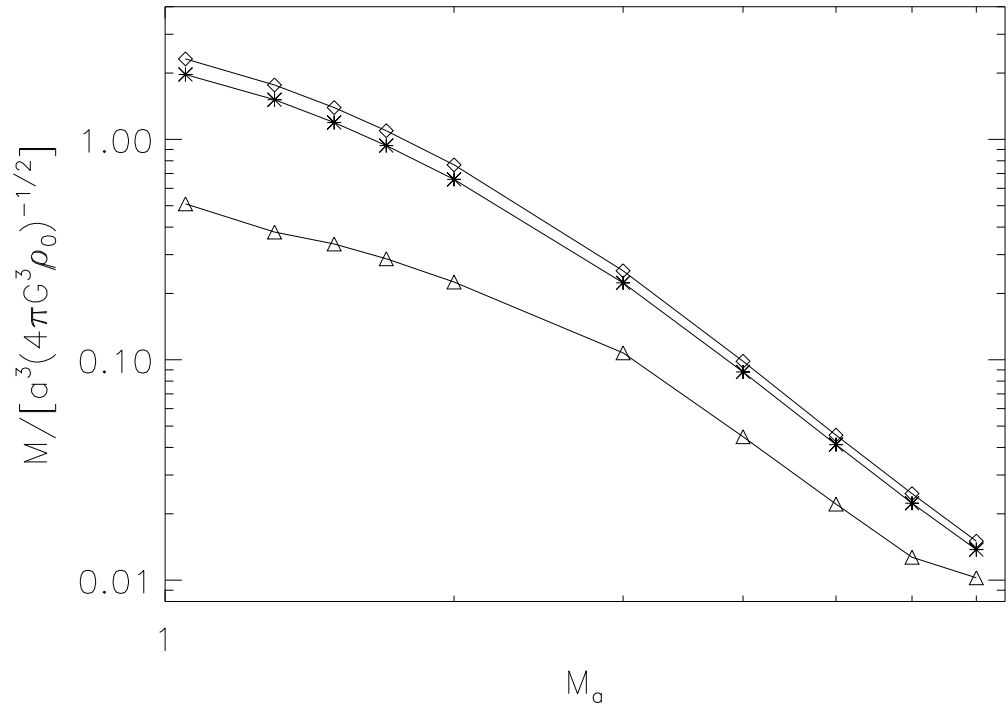}
}
\caption{The protostellar mass at the end of the infall stage (diamonds), 
the core mass inside the shock at the instant $\tau_{coll}$ of protostar 
formation (asterisks), 
and the critical BE sphere mass based on the post-shock density at time
$\tau_{coll}$, all as a function of Mach number $M_a$. The mass unit is given
by equation (\ref{M0_def}).}
\label{fig:core_mass}
\end{figure}

\end{document}